\documentclass{jfm}
\usepackage[dvipsnames]{xcolor} 
\usepackage{graphicx}
\usepackage{tikz}
\usepackage{subcaption}
\usepackage{mathrsfs}
\usepackage{newtxtext}
\usepackage{newtxmath}
\usepackage{natbib}
\usepackage{hyperref}
\usepackage{romannum}
\usepackage{caption}
\hypersetup{
    colorlinks = true,
    urlcolor   = blue,
    citecolor  = black,
}

\def \be {\begin{equation}}
\def \ee {\end{equation}}

\def \bsea{\begin{subeqnarray}} 
\def \esea{\end{subeqnarray}}

\title{Zonal flows driven by libration in rotating spherical shells: the case of periodic characteristic paths}

\author{Xu Chang\aff{1}
\corresp{\email{xu.chang@univ-amu.fr}},
Jiyang He\aff{1,2}, Benjamin Favier\aff{1},
\and St\'ephane  Le Diz\`es\aff{1}}

\affiliation{
	\aff{1}Aix Marseille Univ, CNRS, Centrale Med, IRPHE, Marseille, France
\aff{2}Department of Ocean Science, The Hong Kong University of Science and Technology, Hong Kong, China}
\begin{document}
\pagenumbering{arabic}

\maketitle

\begin{abstract}

	This work investigates the weakly nonlinear dynamics of internal shear layers and the mean zonal flow induced by the longitudinal libration of an inner core within a spherical shell. Building on the work of He et al. (J. Fluid Mech., vol. 939, 2022, A3), which focused on linear dynamics, we adopt a similar setup to explore the nonlinear regime using both asymptotic theory and numerical computations, with Ekman numbers as low as $E=10^{-10}$.	A specific forcing frequency of $\widehat{\omega}=\sqrt{2}\widehat{\Omega}$, where $\widehat{\Omega}$ denotes the rotation rate, is introduced to generate a closed rectangular path of characteristics for the inertial wave beam generated at the critical latitude. Our approach extends previous results by Le Diz\`es (J. Fluid Mech., vol. 899, 2020, A21) and reveals that nonlinear interactions are predominantly localized around regions where the wave beam reflects on the boundary.
	We derive specific scaling laws governing the nonlinear interactions: the width of the interaction region scales as $E^{1/3}$, and the amplitude of the resulting mean zonal flow scales as $E^{1/6}$ in general.
	However, near the rotation axis, where the singularity of the self-similar solution becomes more pronounced, the amplitude exhibits a scaling of $E^{-1/2}$. In addition, our study also examines the nonlinear interactions of beams which are governed by different scaling laws. Through comparison with  numerical results, we validate the theoretical predictions of the asymptotic framework, observing good agreement as the Ekman number decreases.

\end{abstract}

\section{Introduction}\label{sec: intro}

Mechanical forces such as libration, precession, and tides, which arise from gravitational interactions, are essential to generate complex fluid flows within astrophysical and geophysical bodies \citep{lebars2015}.
In the subsurface oceans of celestial bodies such as Enceladus, such forcing initiates dynamical processes that are crucial to understanding the internal structures of these bodies \citep{noir2009,thomas2016,soderlund2024}.
The energy dissipation from libration-driven turbulence may provide heat sources that maintain these subsurface oceans, while libration has been proposed as a potential mechanism driving planetary dynamos \citep{lebars2011,wu2013,reddy2018,wilson2018}.
In the Earth's ocean, the interaction of tidal flows and supercritical topographies is an important source for creating strong concentrated internal wave beams, which play a key role in tidal conversion \citep{smith2003,balmforth2009,echeverri2010}.

In rotating fluids, when the external forcing frequency is less than twice the rotation frequency, smooth inertial modes can be excited within containers where regular inertial modes exist \citep{greenspan1968}.
However, the dynamics becomes more complex in geometries such as spherical shells, where regular inertial modes generally do not exist.
Instead, inertial wave beams are generated at critical latitudes due to oscillatory viscous concentrated boundary layer singularities \citep{kerswell1995,rieutord1997}.
Historical studies have shown that such wave beams can be well described by self-similar solutions \citep{moore1969,thomas1972}.
These solutions have been adapted to more complex cases, such as those involving beam reflections or limit cycles known as attractors \citep{maas1997,rieutord2001,ledizes2017}.
\citet{he2022} expanded the work of \citet{ledizes2017} to a closed domain such as the spherical shell by deriving the asymptotic explicit expression of the linear harmonic velocity as a sum of many self-similar beams.

Transitioning to nonlinear phenomena, the interaction among wave beams introduces complexities, notably when beams intersect or reflect.
Such interactions can lead to the creation of mean flow and second-harmonic corrections.
For example, when beams of the same frequency interact as a result of reflection, they generate significant mean zonal flow and harmonic effects \citep{tabaei2003,tabaei2005,peacock2005}.
These nonlinear processes manifest themselves notably in the formation of mean zonal flows, which have been extensively documented through experimental and numerical studies \citep{tilgner2007,sauret2013,favier2014}.
Previous studies have demonstrated that, in the absence of inertial waves, nonlinear interactions within viscous boundary layers can drive zonal flows. The characteristics of these flows are strongly influenced by the libration amplitude, while remaining largely independent of the Ekman number and depend on the cylindrical variable $r$ only \citep{busse2010,sauret2013}. The dominant contribution to the mean flow arises in the bulk, manifesting as an azimuthal flow that scales as $\epsilon^2$, where $\epsilon$ denotes the small oscillation amplitude.
However, when the libration frequency falls below twice the rotation rate, allowing for the excitation of inertial waves, the resulting interactions become more complex, influencing the structure and behavior of mean zonal flows.
This intricate interplay remains a significant challenge in fluid dynamics research within enclosed domains, where wave reflections and interactions can alter fundamental fluid behaviours \citep{cebron2021}.

Of particular interest to the present study, \citet{tilgner2007} explored the dynamics of mean flow within a rotating spherical shell under the influence of an oscillating tidal mode.
His research revealed that Reynolds stresses are primarily concentrated along wave beam paths, with an intensity peak observed at reflection points and beam intersections. In a subsequent investigation of thin viscous beams reflecting on flat boundaries in rotating and stratified fluids, \citet{ledizes2020} found that when the libration frequency $\omega$ is less than the rotation rate $\Omega$ ($\omega < \Omega$), the reflection process generates both a second-harmonic correction and a mean flow correction. The second-harmonic beam exhibits a larger amplitude compared to the mean flow correction throughout most of the domain, except within the local interaction region.

\citet{ledizes2020} noted that the structure of the wave beam reflections, which have a characteristic width scaling as $E^{1/3}$, maintained self-similarity with a $O(E^{1/6})$ correction, where $E$ is the Ekman number. For scenarios involving purely stratified or rotating fluids, a mean flow correction occurs with an amplitude of $\epsilon^2 E^{-1/6}$, except in cases where the boundary is either horizontal or vertical, which exhibit a localized mean flow correction with a distinct triple-layer structure, including a significant $O(E^{4/9})$ viscous layer.
More recent studies by \citet{lin2021} focused on the numerical analysis of nonlinear mean flow in a spherical shell with libration forcing at the inner core boundary.
They observed that the mean flow correction remains localized at reflection points, scaling approximately as $E^{-1/6}$. Multiple bands were also identified in the bulk flow, directly corresponding to the positions of the reflection points.
However, discrepancies were noted between theoretical predictions and numerical results, particularly regarding slight increases in the mean flow amplitude with the Ekman number.

\begin{figure}
	\centering
	\includegraphics[width=0.95\textwidth]{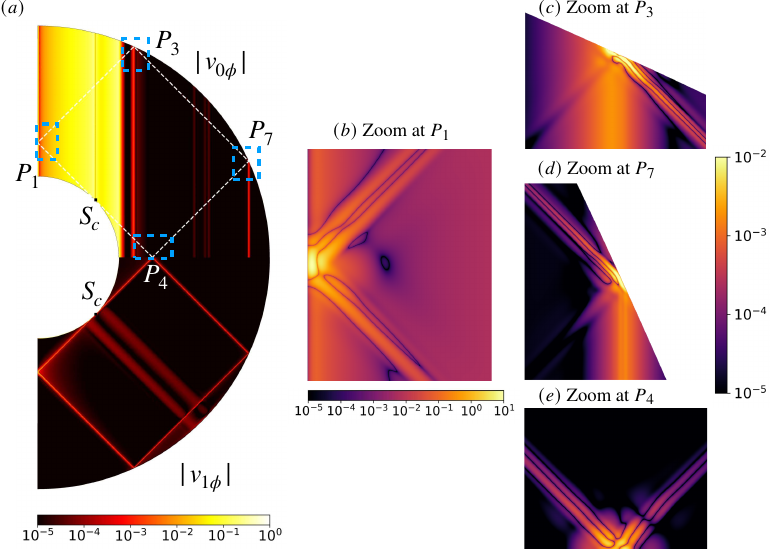}
	\caption{$(a)$ Contours of the amplitudes of the mean flow azimuthal velocity $|v_{0\phi}|$ (upper part) and the linear harmonic solution $|v_{1\phi}|$ (lower part) for $E = 10^{-10}$.  The harmonic solution and the mean flow correction have been normalized by $\varepsilon$ and $\varepsilon^2$, respectively, where $\varepsilon$ is the libration amplitude (see equation~\eqref{eq: velo-asy}). The critical latitude is marked at $S_c$ whose cylindrical coordinates are $(\eta \sqrt{1 - \omega^2/4}, \eta \omega/2)$. The aspect ratio of the spherical shell is $\eta = 0.35$, and the libration frequency of the inner core is $\omega = \sqrt{2}$. The white dashed lines represent the path of characteristics emitted from the critical latitude. $(b)$-$(e)$ Zoomed-in regions of the interaction points located at $P_1$, $P_3$, $P_7$, and $P_4$, respectively.}
	\label{fig: contourmap}
\end{figure}

In this paper, we extend the study to nonlinear mean flow within a rotating spherical shell subjected to libration forcing at the inner core boundary.
Our objective is to use the linear self-similar solutions within the spherical shell to establish a theoretical framework for addressing mean flow corrections. This involves generalizing the previous results of \citet{ledizes2020} and validating the scaling of the mean flow via numerical integration.
For illustrative purposes, the internal shear layers in a spherical shell, induced by the libration of the inner core, along with the corresponding mean flow, are depicted in figure~\ref{fig: contourmap}(a) at a low Ekman number $E=10^{-10}$.
Using direct numerical integration of the linear viscous governing equations, we first compute the linear harmonic velocity field. From this velocity field, we subsequently calculate the mean flow through a pseudo-spectral method. A detailed analysis of the interaction regions reveals that mean flow corrections are significantly larger in those regions than in both the outer region and the bulk bands,
as shown in figures~\ref{fig: contourmap}(b)-(e). These high-amplitude regions are of particular interest, motivating the development of an asymptotic theory to characterize the mean flow correction within these specific regions.

The structure of the paper is as follows: section~\ref{sec: framework} introduces the configuration and settings of the problem. Section~\ref{sec: basic-equation} outlines the fundamental equations. In section~\ref{sec: num-method}, we review the numerical methods employed to solve the linear harmonic governing equations and extend these approaches to address the nonlinear mean flow. Section~\ref{sec: asy-method} revisits the asymptotic theory, examining self-similar solutions and scaling laws. It also summarizes the structure of the  linear harmonic solution for this configuration, providing approximations for this solution close to points where the interactions will be the strongest.
The mean flow correction is addressed in section~\ref{sec: results} and \ref{sec: num-scalings}. We first compare numerical results with the predictions known for the contribution generated from oscillating boundary layers, in \S \ref{sec: bulk}.
We then consider the nonlinear interactions along the periodic beam, especially in the places where the interactions are the strongest, on the axis, and at the reflection points on the boundary (\S \ref{sec: beam}).  Asymptotic expressions are derived and compared to numerical results.
In \S \ref{sec: num-scalings}, we focus on the mean flow band structures that originate from the strong interaction points. The scaling  and form of the solution in these bands are analysed numerically
using the asymptotic prescriptions that can be derived for these structures (appendix \S \ref{app:band}).
Finally, the paper concludes with section~\ref{sec: conclusion}, which summarizes the key findings and discusses their consequences for other configurations.

\section{Framework}\label{sec: framework}
\begin{figure}
	\centering
	\includegraphics[width=0.95\textwidth]{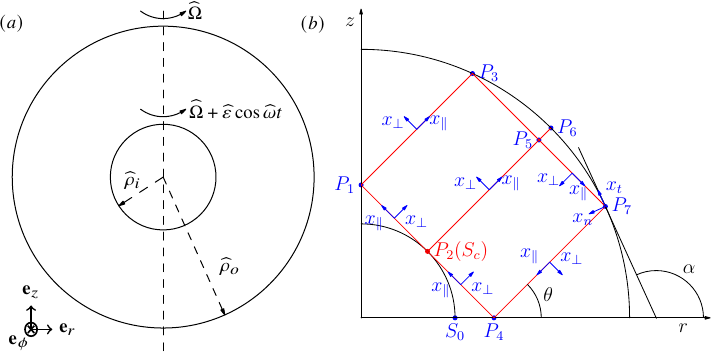}
	\caption{Sketch of the problem: $(a)$ the outer radius of the concentric shell is $\widehat{\rho}_o$ and rotates at the angular velocity of $\widehat{\Omega}$, the inner radius is $\widehat{\rho}_i$ and rotates at an angular velocity of $\widehat{\Omega}+\widehat{\varepsilon} \cos \widehat{\omega} t$ with the amplitude $\widehat{\varepsilon}$ and frequency $\widehat{\omega}$ of the libration; $(b)$ A local coordinate system is defined in the northward direction. The red lines represent the critical lines, while $x_{\|}$ and $x_{\perp}$ are the local coordinates. The southward direction is the inverse of the northward direction.}
	\label{fig: configure}
\end{figure}

In this paper, we study the dynamics of an incompressible fluid with a constant kinematic viscosity $\nu$, which fills a spherical shell and rotates around the axis $\mathbf{e}_z$ at a uniform rate $\widehat{\Omega}$.
The flow is further subjected to the libration of the inner core, as depicted in figure~\ref{fig: configure}(a).
The inner core librates with a small amplitude $\widehat{\varepsilon}$ and at a frequency $\widehat{\omega}=\sqrt{2}\widehat{\Omega}$, resulting in an angular velocity of $\widehat{\varepsilon} \cos(\widehat{\omega} t)$ relative to the rotating system.
Following the configuration of \citet{he2022}, the radii of the inner and outer spherical cores are $\widehat{\rho}_i$ and $\widehat{\rho}_o$, respectively, with the aspect ratio $\eta = \widehat{\rho}_o/\widehat{\rho}_i= 0.35$ to generate an Earth-like ratio.
Time and space are non-dimensionalized using the inverse rotation rate $1/\widehat{\Omega}$ and the outer sphere radius $\widehat{\rho}_o$ of the spherical shell.
The non-dimensional inner core radii is then $\eta$. The non-dimensional angular velocity of the outer and inner core are $1$ and $1+\varepsilon \cos \omega t$ respectively, with libration amplitude $\varepsilon = \widehat{\varepsilon}/\widehat{\Omega}$ and frequency $\omega=\widehat{\omega}/\widehat{\Omega}$.

\subsection{Basic equations}\label{sec: basic-equation}

In the rotating frame, the velocity $\mathbf{V}=(v_r,v_{\phi},v_z)$, written in cylindrical coordinates, and pressure $P$ are governed by the following equations
\begin{subequations}
	\begin{align}
		\frac{ \partial \mathbf{V} }{ \partial t } + (\mathbf{V}\cdot \nabla)\mathbf{V} + 2\mathbf{e}_{z}\times \mathbf{V} & =-\nabla P+E\nabla^{2}\mathbf{V} \label{eq: basic-eqn} \ , \\
		\nabla \cdot \mathbf{V}                                                                                            & = 0 \ ,
	\end{align}
\end{subequations}
where $E$ is the Ekman number, which is defined as $E = \nu/\widehat{\Omega} \widehat{\rho}_o^{2}$.

Contrarily to \citet{he2022}, which exclusively addressed the linear aspects by omitting the advection term in equation~\eqref{eq: basic-eqn}, this study expands into the nonlinear regime. We consider the following asymptotic expansion of the velocity $\mathbf{V}$  and pressure $P$:
\begin{equation}
	(\mathbf{V},P) = \varepsilon (\mathbf{v}_1, p_1)e^{-\mathrm{i}\omega t}+ \varepsilon^2\left[(\mathbf{v}_0,p_0)+(\mathbf{v}_2,p_2)e^{-2\mathrm{i}\omega t} \right]+ \mathrm{c}.\mathrm{c},
	\label{eq: velo-asy}
\end{equation}
where $\mathrm{c}.\mathrm{c}$ denotes the complex conjugate, $(\mathbf{v}_1,p_1)$ is the linear harmonic, $(\mathbf{v}_0,p_0)$ is the mean flow correction, and $(\mathbf{v}_2,p_2)$ is the secondary harmonic velocity.
In this paper, we focus solely on the mean flow correction.

Three steps are required to compute the mean flow correction.
First, similarly to \cite{he2022}, the linear response is found by solving
\begin{subequations}
	\begin{align}
		-\mathrm{i}\omega \mathbf{v}_1+2\mathbf{e}_{z} \times \mathbf{v}_1 & =-\nabla p_1+E\nabla^{2} \mathbf{v}_1 \ , \label{eq: linear-governing-equation} \\
		\nabla \cdot \mathbf{v}_1                                          & =0 \ ,                                                                          \\
		\text{Inner boundary condition:} \quad \mathbf{v}_1                & = r\mathbf{e}_{\phi} \ ,                                                        \\
		\text{Outer boundary condition:} \quad \mathbf{v}_1                & = \mathbf{0} \ .
	\end{align}\label{eq: linear-governing}
\end{subequations}
Once the linear harmonic velocity $\mathbf{v}_1$ is obtained, the nonlinear terms corresponding to its self-interaction, can be calculated at leading order.
The resulting Reynolds stresses comprise two parts: the steady forcing and the second harmonic oscillating at $2\omega$.
Throughout this paper, we refer to the divergence of the Reynolds stress tensor (which provides the mean flow forcing) simply as the ``Reynolds stress''.
As the present study focuses on the mean flow, we only consider the steady Reynolds stress $\mathcal{N}_{0}$
\begin{equation}
	\mathbf{\mathcal{N}}_{0}=\mathbf{v}_1\cdot \nabla \mathbf{v}_{1}^{*} + \mathrm{c}.\mathrm{c} \ , \label{eq: steady-rss}
\end{equation}
where $^*$ denotes complex conjugation.
The mean flow response $\mathbf{v}_0$ corresponds to a perturbation with a zero frequency $\omega=0$ and is governed by a linear forced system:
\begin{subequations}
	\begin{align}
		2\mathbf{e}_{z} \times \mathbf{v}_{0}+\nabla p_{0}            & =-\mathbf{\mathcal{N}}_{0}+E \nabla^{2} \mathbf{v}_{0} \ , \\
		\nabla \cdot \mathbf{v}_0                                     & =0 \ ,                                                     \\
		\text{Inner and outer boundary condition:} \quad \mathbf{v}_0 & =\mathbf{0} \ .
	\end{align}
	\label{eq: nonlinear-governing}
\end{subequations}
where the forcing $\mathbf{\mathcal{N}}_0$ arises from the nonlinear self-interaction of the linear harmonic $v_1$.

\subsection{Numerical approach}\label{sec: num-method}

To solve the equations outlined in section~\ref{sec: basic-equation} and validate our theoretical predictions through asymptotic analysis, we have employed high-precision spectral methods for the numerical integrations.
These methods, previously used in \cite{rieutord1997} and \citet{he2022} to solve the linear governing equation \eqref{eq: linear-governing}, have enabled us to capture the linear harmonic velocity profiles $\mathbf{v}_{1}$ displayed in figure~\ref{fig: contourmap}(a).
We have looked for the solution of the linear governing equations~\eqref{eq: linear-governing-equation} expressed in vorticity form as
\begin{equation}
	-\mathrm{i}\omega \nabla \times \mathbf{v}_1+2\nabla \times(\mathbf{e}_z\times \mathbf{v}_1)=E\nabla \times(\nabla^{2}\mathbf{v}_1) \ .
	\label{eq: vorticity-equation}
\end{equation}
For this three-dimensional problem, we use spherical coordinates $(\rho, \theta, \phi)$ representing radial, polar, and azimuthal directions, respectively. The velocity fields are expanded using spherical harmonics in the polar and azimuthal directions and Chebyshev polynomials in the radial direction:
\begin{equation}
	\mathbf{v}_{1}=\sum^{+\infty}_{l=0}\sum^{+l}_{m=-l}u_{m}^{l}(\rho)\mathbf{R}_{l}^{m}+v^{l}_{m}(\rho)\mathbf{S}_{l}^{m}+w_{m}^{l}(\rho)\mathbf{T}_{l}^{m} \ ,
\end{equation}
where,
\begin{equation}
	\mathbf{R}_{l}^{m}=Y_{l}^{m}(\theta,\phi)\mathbf{e}_{\rho},\quad \mathbf{S}_{l}^{m}=\nabla Y_{l}^{m},\quad \mathbf{T}_{l}^{m}=\nabla \times \mathbf{R}_{l}^{m} \ .
\end{equation}
Projecting the vorticity equation~\eqref{eq: vorticity-equation} onto this basis allows us to solve the linear system using a block tridiagonal algorithm. In our cases, we are assuming an axisymmetric fluid response along with the no-slip boundary conditions on the spherical shell.
Detailed descriptions of the numerical methods and the construction of the system are available in \citet{he2022} and \citet{he2023} for readers interested in further details.

After obtaining the linear harmonic response numerically, we apply a pseudo-spectral method to calculate the steady Reynolds stress.
Subsequently, we solve the governing equations~\eqref{eq: nonlinear-governing} to determine the mean flow velocity, $\mathbf{v}_0$, shown in figure~\ref{fig: contourmap}(a).
The nonlinear vorticity equation for the mean zonal flow is expressed as
\begin{equation}
	2\nabla \times(\mathbf{e}_z\times \mathbf{v}_0)=-\nabla \times(\mathbf{v}_1\cdot \nabla \mathbf{v}_1^{*}+\mathrm{c}.\mathrm{c}) +E \nabla\times \nabla^{2}\mathbf{v}_{0}\ .
	\label{eq: nonlinear-vorticity-equation}
\end{equation}
Details on the pseudo-spectral numerical workflow are provided in appendix~\ref{sec: app-num} and illustrated in figure~\ref{fig: num-map}.
We have also investigated potential aliasing issues in this nonlinear setup and  found no significant impact on the results, regardless of whether dealiasing was applied or not.
Consequently, all the results discussed in the present paper have not been de-aliased.
The convergence of our simulations is further demonstrated in appendix~\ref{sec: app-num} and the numerical resolutions for different Ekman numbers are documented in table~\ref{tab: resolution} of this appendix.
It is important to note that the resolutions required for the nonlinear simulations exceed those used in the purely linear analyses,  to ensure similar convergence properties in both cases.

\section{Asymptotic description of the harmonic solution}\label{sec: asy-method}

As first demonstrated by \citet{moore1969}, the propagation and viscous smoothing of a localized singularity can be described, in the limit of small Ekman numbers, by a self-similar solution.
This viscous smoothing leads to the emergence of a self-similar expression for the primary components of the wave beam velocity.
Note that the concentrated beams that originate from the critical latitude are linked to an inviscid singularity along the critical ray, as identified by \citet{ledizes2024}.

In the scenario under consideration, it is postulated that concentrated beam rays emanate from the critical latitude singularity where the boundary is locally tangent to the direction of inertial wave propagation, denoted as $S_c(r,z)=(\eta\sqrt{1-\omega^2/4},\eta\omega/2)$, and propagate in two distinct directions: clockwise, designated as the northward direction, following the path $S_c \to P_1 \to P_3 \to P_7 \to P_4 \to S_c$, and counterclockwise, referred to as the southward direction.
As shown in figure~\ref{fig: configure}(b), $P_1$ is the point on the characteristic path crossing the rotation axis, $P_3$ and $P_7$ are the reflection points on the outer boundary while $P_4$ is on the equator where two beams are crossing. A local coordinate system $(x_{\parallel}, x_{\perp})$ is introduced to locally describe the asymptotic structure of the beam, where $x_{\parallel}$ represents the distance from the source located at the critical latitude point $S_c$, and $x_{\perp}$ denotes the distance measured perpendicularly to the direction of beam propagation.
For simplicity, it is assumed that the orientation of propagation does not change sign.
When considering southward propagation, the direction of the corresponding local coordinate system is reversed.

In the following subsections, we provide expressions for the harmonic solution in the various regions where the mean flow correction becomes significant. We first review the results obtained by \citet{he2022} on the characteristic path, then present asymptotic expressions near the various interaction regions.

\subsection{Self-similar solution and scaling}\label{sec: self-similar}

For the harmonic solution, the localized concentrated beams travel on the beam ray with a constant width of order $E^{1/3}$. The self-similar beam solution provided by \citet{moore1969} was used by \citet{ledizes2017} and \cite{he2022} to describe at leading order the velocity $v_{1\parallel}$ along the direction of the beam propagation,
\begin{equation}
	v_{1\parallel}\sim \frac{1}{\sqrt{ r }}C_{0}H_{m}(x_{\parallel},x_{\perp})=\frac{1}{\sqrt{ r }}C_{0}\left( \frac{x_{\parallel}}{2\sin\theta} \right)^{-m/3}h_{m}(\zeta) \ .\label{eq: self-similar-expression}
\end{equation}
The special function $h_m(\zeta)$ is,
\begin{equation}
	h_m(\zeta)=\frac{e^{-\mathrm{i}m\pi/2}}{(m-1)!}\int_0^{+\infty}e^{\mathrm{i}p\zeta-p^3}p^{m-1}dp \ ,
	\label{eq: hm}
\end{equation}
with the similarity variable,
\begin{equation}
	\zeta=\frac{x_{\perp}}{E^{1/3}}\left( \frac{2\sin\theta}{x_{\parallel}} \right)^{1/3} \ .\label{eq: zeta}
\end{equation}
$\theta$ is the angle between the inertial wave propagation direction and the equatorial plane, where $\theta = \pi/4$ enables the formation of a simple closed wave circuit in our configuration (see figure~\ref{fig: configure}(b)).
This angle also defines the inclination of the internal shear layers relative to the equatorial plane, with the libration frequency $\omega = 2\cos\theta$.
The index $m$ of the function $h_m(\zeta)$ characterizes the nature of the singularity that leads to the formation of the beam \citep{moore1969, thomas1972}, while the complex amplitude $C_0$ represents the amplitude of the singularity.
A brief review of different values of the index $m$ for various configurations is provided by \citet{ledizes2020}.
In particular, for any axisymmetric convex librating object, such as the inner sphere considered in the present study, \citet{ledizes2017} determined the values $m = 5/4$ and $C_0$ by matching the similarity solution with the boundary layer solution around the critical latitude leading to
\begin{subequations}
	\begin{align}
		C_0^N & = E^{1/12}\tilde{C}_{0}^{N}=E^{1/12}\frac{e^{\mathrm{i}\pi/2}}{8(2\sin\theta)^{3/4}}, \quad \text{northward ray} \ , \\
		C_0^S & =E^{1/12}\tilde{C}_{0}^{S}=E^{1/12}\frac{e^{\mathrm{i}3\pi/4}}{8(2\sin\theta)^{3/4}}, \quad \text{southward ray} \ .
	\end{align}\label{eq: ledizes-amplitude}
\end{subequations}
Expression \eqref{eq: self-similar-expression} describes the main velocity component. In the $(r,z)$ plane, there also exists a component along ${\bf e}_\perp$ which is $O(E^{1/3})$ smaller.

As shown by equations~\eqref{eq: ledizes-amplitude}, the velocity amplitude scales as $E^{1/12}$  in the asymptotic theory.
Internal shear layers within a spherical shell were observed to preserve their self-similar structure upon reflecting on boundaries, undergoing either contraction or expansion during this process.
Particular attention was given to how these layers propagate along a periodic rectangular path at specific frequencies of the libration forcing \citep{he2022}. Additionally, reflections on the rotation axis were found to induce a phase shift, facilitating the convergence of the summation describing the beam superpositions.
The final expressions for the velocity components were derived from these self-similar solutions, resulting in the following compact form in the $(r,z)$ plane:
\begin{equation}
	{\bf v}_{1_{2D}}\sim \frac{E^{1/12}}{\sqrt{r}}(\tilde{C}_{0}^{N}G_{m}(x_{\perp}^{N},x_{\parallel}^{N},L^{N}) {\bf e}_{\parallel}^N+\tilde{C}_{0}^{S}G_{m}(x_{\perp}^{S},x_{\parallel}^{S},L^{S}){\bf e}_{\parallel}^S) \ .
	\label{eq: final-linear-solution}
\end{equation}
where the new function $G_m$ is given by
\begin{subequations}
	\begin{align}
		 & G_{m}(x_{\perp},x_{\|},L)=\left( \frac{x_{\parallel}}{2\sin\theta} \right)^{-m/3}g_{m}(\zeta,L/x_{\parallel}) \ ,                                               \\
		 & g_{m}(\zeta,L/x_{\parallel})=\frac{e^{-\mathrm{i}m\pi/2}}{(m-1)!}\int_0^\infty\frac{e^{\mathrm{i}p\zeta-p^3}p^{m-1}}{1-\mathrm{i}e^{-p^3L/x_{\parallel}}}dp \ .
	\end{align}
	\label{eq: gm}
\end{subequations}
Here, $L$ is the cumulative propagation distance along the rectangular characteristic path for one cycle (explicit form in appendix~\ref{sec: app-parameters}, equation~\eqref{eq: L}).

Compared to the function $h_m$ in \eqref{eq: hm}, the denominator of $g_m$ comes from the beam summation and the ``$\mathrm{i}$" factor from the phase shift induced by the reflection on the axis. Note that ${\bf e}_{\parallel}^S= -{\bf e}_{\parallel}^N$.

Expression \eqref{eq: final-linear-solution} describes the solution on the rectangular circuit $P_1P_3P_7P_4$.
As explained in  \cite{he2022}, a weaker secondary beam is also present between $S_c$ and $P_6$ owing to the reflection on the inner core.
The width scaling of this secondary beam is $E^{1/6}$, indicating a relatively larger transverse scale on which viscous diffusion becomes negligible. Consequently, the parallel velocity component $v_{1\parallel}$ of the beam remains constant throughout its path. Furthermore, the amplitude of the beam scales as $E^{1/4}$, which is significantly smaller than the amplitude of the solution on the main beam.  The velocity profile along the beam is asymptotically described by
\begin{equation}
	{\bf v}_{1_{2D}} \sim \frac{E^{1/4}}{\sqrt{r}}(  \tilde{C}_0^{N}\breve{F}_{m}(\breve{x}_{\perp}^N,\lambda^{N}) {\bf e}_{\parallel}^N + \tilde{C}_0^{S}\breve{F}_{m}(\breve{x}_{\perp}^S,\lambda^{S}) {\bf e}_{\parallel}^S)
	\label{eq: linear-weak}
\end{equation}
where
\begin{equation}
	\breve{F}_m(\breve{x}_{\perp},\lambda)  =\frac{\breve{x}_\perp}{\sqrt{2}}\frac{e^{-\mathrm{i}m\pi/2}}{(m-1)!}\int_0^\infty e^{\mathrm{i}p\lambda(\breve{x}_{\perp})^2}\frac{e^{-p^3L/\sqrt{2}}p^{m-1}}{1-\mathrm{i}e^{-p^3L/\sqrt{2}}}dp \ ,
\end{equation}
with  $\lambda^{N}=-\sqrt{2}/4$, $\lambda^{S}=\sqrt{2}/4$ and $\breve{x}_{\perp}=x_\perp/E^{1/6}$.

A detailed comparison between this asymptotic solution and numerical solutions of the linear harmonic viscous problem can be found in \cite{he2022}.

The azimuthal velocity, $v_{1\phi}$, can be readily derived from $v_{1\|}$ using
\begin{equation}
	v_{1\phi}\sim\pm \mathrm{i}v_{1\|}\label{eq: v1phi} \ ,
\end{equation}
where the positive sign (resp. negative sign) corresponds to an obtuse angle (resp. acute angle) between $\mathbf{e}_{\parallel}$ and $\mathbf{e}_r$.
This equation also implies, for $\theta=\pi/4$, that
\be
v_{1\phi}\sim -\sqrt{2}\mathrm{i} v_{1r}.
\label{eq: v1phi2}
\ee
The velocity field defined by \eqref{eq: final-linear-solution} and \eqref{eq: linear-weak} also satisfies, for $\theta=\pi/4$, the remarkable property
\be
\frac{\partial v_{1r} }{\partial z} \sim - \frac{\partial v_{1z} }{\partial r} .
\label{eq: v1r}
\ee
The symbol $\sim$ in equations (\ref{eq: v1phi}), (\ref{eq: v1phi2}) and (\ref{eq: v1r}) has to be understood as denoting equality up to corrections of order $E^{1/3}$.
We shall see below the importance of these equations  for simplifying the expression of the Reynolds stress.

For detailed information about the local coordinate system relationships of northward and southward ray propagation paths, as well as more details about the different beam amplitudes, see table~\ref{tab: north-path} and table~\ref{tab: south-path} in appendix~\ref{sec: app-parameters}.
In the following, we shall often keep the parameter $m$ unprescribed to show the dependency of the various expressions with respect to this parameter, but one has to keep in mind
that in the present analysis $m$ and $\theta$ are actually fixed to  $m=5/4$ and $\theta=\pi/4$.

\begin{figure}
	\centering
	\includegraphics[width=0.7\textwidth]{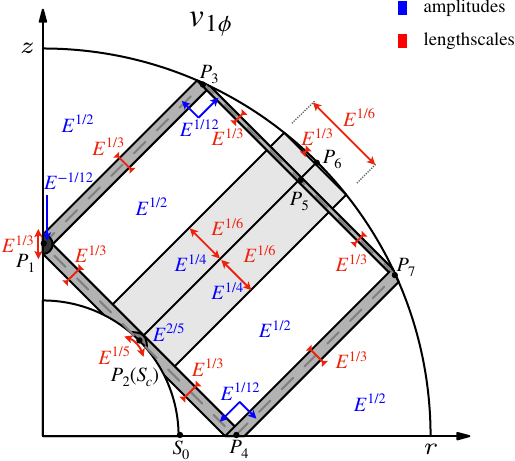}
	\caption{Ekman scalings of the azimuthal velocity $v_{1\phi}$ of the linear harmonic solution, reproduced from figure 4 in \citet{he2022}, while the amplitudes and the lengthscales are shown in blue and red respectively.}
	\label{fig: linear-scaling}
\end{figure}

Figure~\ref{fig: linear-scaling} illustrates the scaling behaviour derived from the linear part of the asymptotic analysis.
In this figure, the beam width scaling are shown in red, while the amplitude scaling are depicted in blue.
In particular, the main beam is characterized by a beam width scaling like $E^{1/3}$ and an amplitude scaling like $E^{1/12}$.
The corresponding scaling for the nonlinear mean flow will be discussed in the results section~\ref{sec: num-scalings}.

\subsection{Harmonic solution near $P_3$, $P_4$ and $P_7$}
\label{sec: reflect}

The beams reflect on the outer boundary at the interaction points $P_3$ and $P_7$ while two beams intersect on the equatorial plane at the point $P_4$, see figure~\ref{fig: configure}(b).
For describing the solution near those points, it is convenient to introduce the local variables
\be
\tilde{r}_\beta= \frac{r-r_{P_\beta}}{E^{1/3}} ~,~~\tilde{z}_\beta= \frac{z-z_{P_\beta}}{E^{1/3}},
\ee
and express the variables $x_\perp$ and $x_\parallel$ of the beam in terms of these local variables.
Both northward and southward beams exhibit two parts, an incident beam, corresponding  to the beam reaching $P_ \beta$ and a reflected beam, corresponding
to the beam leaving $P_\beta$. 
Note that the point $P_4$ is not on a boundary, but the symmetry of the solution with respect to the equator has the effect of reflecting the beam on the horizontal plane.
We shall use the compact notation $\tilde{B}_{\beta}^{Ni}$, $\tilde{B}_{\beta}^{Nr}$, $\tilde{B}_{\beta}^{Si}$, $\tilde{B}_{\beta}^{Sr}$ to denote the four contributions present near $P_\beta$.
The northward incident beam reaching $P_\beta$ is for instance given by
\be
\tilde{B}_{\beta}^{Ni}(\tilde{r}_\beta, \tilde{z}_\beta)  =\tilde{C}_{0\beta}^{Ni}G_{m}\left(x_{\perp \beta}^{Ni}, x_{\parallel \beta}^{Ni}, L_\beta^{Ni}\right) \ , \label{eq: Bm}
\ee
where the parameters $\tilde{C}_{0\beta}^{Ni}$ and $L_\beta^{Ni}$ and the variables $x_{\perp \beta}^{Ni}$ and $ x_{\parallel \beta}^{Ni}$ can be obtained using table~\ref{tab: north-path}
of Appendix~\ref{sec: app-parameters}. For the southward beam, these quantities are given in table~\ref{tab: south-path}.

In each local region around $P_3$,  $P_4$ and $P_7$, we can write the velocity field as
\begin{equation}
	\mathbf{v}_{1}\sim \frac{1}{\sqrt{r}}E^{1/12} \mathbf{\tilde{v}}_{1} \ ,
\end{equation}
with
\bsea
&&\tilde{v}_{1r}     =2^{-1/2}\tilde{B}_3^{+} \ , ~~
\tilde{v}_{1\phi}  =-\mathrm{i}\tilde{B}_3^{+} \ ,~~
\tilde{v}_{1z}     =2^{-1/2}\tilde{B}_3^{-} \ ,  ~~ {\rm for }~~P_3, \\
&&\tilde{v}_{1r}     =-2^{-1/2}\tilde{B}_4^{+} \ , ~~
\tilde{v}_{1\phi}  =\mathrm{i}\tilde{B}_4^{+} \ ,~~
\tilde{v}_{1z}     =-2^{-1/2}\tilde{B}_4^{-} \ , ~~ {\rm for }~~P_4,  \\
&&\tilde{v}_{1r}    =2^{-1/2}\tilde{B}_7^{-} \ ,~~
\tilde{v}_{1\phi}  =-\mathrm{i}\tilde{B}_7^{-} \ , ~~
\tilde{v}_{1z}     =-2^{-1/2}\tilde{B}_7^{+} \ ,   ~~ {\rm for }~~P_7,
\label{eq: v1-p7}
\esea
where
\bsea
\tilde{B}_\beta^{+}(\tilde{r}_\beta,\tilde{z}_\beta) = ( \tilde{B}_{\beta}^{Ni}-\tilde{B}_{\beta}^{Sr}) +(\tilde{B}_{\beta}^{Nr}-\tilde{B}_{\beta}^{Si}), \\
\tilde{B}_\beta^{-}(\tilde{r}_\beta,\tilde{z}_\beta) =  (\tilde{B}_{\beta}^{Ni}-\tilde{B}_{\beta}^{Sr})-(\tilde{B}_{\beta}^{Nr}-\tilde{B}_{\beta}^{Si}).
\esea

There is a relation between incident and reflected beams: they are such that the normal velocity of the sum of these two contributions vanish on the boundary.
This condition can be written, at the three points $P_3$, $P_4$ and $P_7$, as
\bsea
&&\tilde{B}^{Ni}_{3}(\tilde{x}_n=0) = K \tilde{B}^{Nr}_{3}(\tilde{x}_n=0)  ~~, ~~~
\tilde{B}^{Si}_{3}(\tilde{x}_n=0) = (1/K)  \tilde{B}^{Sr}_{3}(\tilde{x}_n=0) , \\
&&\tilde{B}^{Ni}_{4}(\tilde{x}_n=0) =  \tilde{B}^{Nr}_{4}(\tilde{x}_n=0)  ~~, ~~~
\tilde{B}^{Si}_{4}(\tilde{x}_n=0) =   \tilde{B}^{Sr}_{4}(\tilde{x}_n=0) , \\
&&\tilde{B}^{Ni}_{7}(\tilde{x}_n=0) = (1/K) \tilde{B}^{Nr}_{7}(\tilde{x}_n=0)  ~~, ~~~
\tilde{B}^{Si}_{7}(\tilde{x}_n=0) = K  \tilde{B}^{Sr}_{7}(\tilde{x}_n=0) ,
\label{eq:Bxn=0}
\esea
where $\tilde{x}_n=0$ denotes the boundary, and
\begin{equation}
	K =\frac{\sin(\alpha+\pi/4)}{\sin(\alpha-\pi/4)}
	\label{exp:K}
\end{equation}
the contraction factor of the northward beam at $P_3$.

However the tangential velocity does not vanish on the boundary.
Consequently, in the local region $P_3$ and $P_7$ where a real boundary is present, a viscous boundary layer is expected to form.
This boundary layer was studied in \citet{ledizes2020}. It was shown to induce a $O(E^{1/6})$ viscous correction to the reflected beam, which is further discussed below in section~\ref{sec:higher}.

\subsection{Harmonic solution near $P_5$ and $P_6$}
\label{sec: linear-p5}

The weak beam propagating between $S_c$ and $P_6$ interacts with the main beam propagating between $P_3$ and $P_7$ at the point $P_5$.
In this interaction region which is $O(E^{1/3})$ large and $O(E^{1/6})$ long in the direction aligned with the main beam,
the harmonic solution exhibits a particular approximation that can be written as
\begin{equation}
	\mathbf{v}_{1} \sim \frac{1}{\sqrt{r}}(E^{1/4}\mathbf{\breve{v}}_{1}+E^{1 /12}\mathbf{\tilde{v}}_{1})\ ,
\end{equation}
with
\bsea
&&	\breve{v}_{1r} = \frac{1}{\sqrt{2}}(\breve{B}^{N}-\breve{B}^{S}) , ~~ 	\breve{v}_{1\phi} = -\mathrm{i}(\breve{B}^{N}-\breve{B}^{S}), ~~  \breve{v}_{1z} = \frac{1}{\sqrt{2}}(\breve{B}^{N}-\breve{B}^{S}), \\
&& \tilde{v}_{1r} = \frac{1}{\sqrt{2}}(\tilde{B}_5^{N}-\tilde{B}_5^{S}),~~
\tilde{v}_{1\phi} = -\mathrm{i}(\tilde{B}_5^{N}-\tilde{B}_5^{S}),~~ \tilde{v}_{1z} = -\frac{1}{\sqrt{2}}(\tilde{B}_5^{N}-\tilde{B}_5^{S})\ ,
\esea
where the beam structures are given for example for the northward beams by
\begin{subequations}
	\begin{align}
		 & \breve{B}^{N} = \tilde{C}_0^{N}\breve{F}_m(\breve{x}_{\perp }^{N},\lambda^{N}) \ ,               \\
		 & \tilde{B}_5^{N}= \tilde{C}_{05}^{N}G_{m}(\tilde{x}_{\perp 5 }^{N},x_{\parallel 5}^{N},L_5^N) \ ,
	\end{align}
\end{subequations}

A similar expression is obtained for $P_6$, upon changing  $\tilde{B}_5^N$ by $\tilde{B}_6^S$ and, $\tilde{B}_5^S$ by $\tilde{B}_6^N$.
However, $P_6$ being on the boundary, a viscous correction is expected. Because $P_6$ is a critical point, this viscous correction is larger than that of $P_3$ and $P_7$. It is expected to be of order $E^{1/12}$ as shown in \citet{ledizes2024}.
No such correction is generated at $P_5$.

\subsection{Harmonic solution near $P_1$}\label{sec: linear-p1}

The point $P_1$ where the critical beam reaches the axis is peculiar.  At that point, the approximation ~\eqref{eq: final-linear-solution} exhibits a singularity, which means that another
approximation must be used in the local region near $P_1$.
In that region, the velocity and pressure should be expressed using the Hankel transform, as discussed by \citet{ledizes2017} (equation (A1) in appendix) and \citet{he2022} (equation (4.1) in their paper).

In \citet{ledizes2015}, the solution was derived for an open-domain configuration, focusing on the local region of beam reflection along the axis. The theory was extended to a closed domain with periodic characteristics in \citet{he2022}.   Introducing the local variables $\tilde{r}_1 = r/E^{1/3}$ and $\tilde{z}_1 = (z - \sqrt{2}\eta ) / E^{1/3}$, they showed that the linear harmonic velocity in $P_1$ region can be written as
\begin{equation}
	\mathbf{v}_{1} \sim E^{-1 / 12}\mathbf{\hat{v}}_{1} \ ,
\end{equation}
where \begin{subequations}
	\begin{align}
		\hat{v}_{1r}    & =  \frac{\mathrm{i}}{\sqrt{2}}\left( Q^{N}_{1}+Q^{S}_{1} \right) \ , \\
		\hat{v}_{1\phi} & =Q^{N}_{1}+Q^{S}_{1} \ ,                                             \\
		\hat{v}_{1z}    & = \frac{1}{\sqrt{2}}\left( -Q^{N}_{0}+Q^{S}_{0}  \right) \ .
	\end{align}\label{eq: p1-linear-velo}
\end{subequations}
The functions  $Q^{N}_l(\tilde{r}_1, \tilde{z}_1)$  and $Q^{S}_l(\tilde{r}_1, \tilde{z}_1)$ are defined, for $l=0,1$,  by
\begin{subequations}
	\begin{align}
		 & Q^{N}_{l}(\tilde{r}_1,\tilde{z}_1)=\int _{0}^{\infty}\hat{V}^{N}(\tilde{k})J_{l}(\tilde{k}\tilde{r}_1) e^{ \mathrm{i}\tilde{k}\tilde{z}_1 }\, d\tilde{k} \ ,  \\
		 & Q^{S}_{l}(\tilde{r}_1,\tilde{z}_1)=\int _{0}^{\infty}\hat{V}^{S}(\tilde{k})J_{l}(\tilde{k}\tilde{r}_1) e^{ -\mathrm{i}\tilde{k}\tilde{z}_1 }\, d\tilde{k} \ ,
	\end{align}\label{eq: asy-linear-p1}
\end{subequations}
where $J_l$ are  Bessel functions of the first kind. The superscripts $N$ and $S$
designate  northward and southward beam respectively.

The functions $\hat{V}^{N}$ and $\hat{V}^{S}$ are obtained by matching the Hankel transform expression to the  solution \eqref{eq: final-linear-solution} valid away from $P_1$. As shown \citet{he2022}, this leads to the following expressions:
\begin{align}
	\hat{V}^{N} & =2^{m/2-3+1/8}\sqrt{ \pi }e^{-\mathrm{i}\pi/4 } \frac{e^{ -\mathrm{i}m\pi/2 }}{(m-1)!} \frac{\tilde{k}^{m-1/2}e^{ -2x_{\parallel}^{N}\tilde{k}^{3} }}{1-\mathrm{i}e^{ -2L\tilde{k}^{3} }}, \\
	\hat{V}^{S} & =2^{m/2-3+1/8}\sqrt{ \pi }\frac{e^{ -\mathrm{i}m\pi/2 }}{(m-1)!} \frac{\tilde{k}^{m-1/2}e^{ -2x_{\parallel}^{S}\tilde{k}^{3} }}{1-\mathrm{i}e^{ -2L\tilde{k}^{3} }},
	\label{eq: p1-coeff}
\end{align}
where we have used expressions \eqref{eq: ledizes-amplitude} for $C_0^{N}$ and $C_0^S$ to simplify the expressions given in \citet{he2022}, eq. (4.8a,b).

Note the scaling of the amplitude in $E^{-1/12}$ and of the region size in $E^{1 / 3}$, which has been illustrated in  figure~\ref{fig: linear-scaling}.
The methodology developed for the $P_1$ region can be generalized to any beam interaction on the rotation axis. Note in particular that the functions $\hat{V}^N$ and  $\hat{V}^S$ in our current analysis contain a term in $1 - \mathrm{i} e^{-2Lk^3}$ in the denominator, which arises from the summation of infinitely many beam contributions owing to the periodicity of the critical path.  For a simple beam interaction without summation,
no such term in the denominator is present \cite[see][]{ledizes2015}.

\subsection{Higher-order corrections to the harmonic solution}
\label{sec:higher}

In the previous sections, we have provided the leading-order approximation of the harmonic response in the interaction regions.
These expressions are based on the similarity solution \eqref{eq: self-similar-expression} obtained by \cite{moore1969}, which is known to be valid in an open domain up to $O(E^{1/3})$ corrections induced by variations along the beam.

Larger viscous corrections are created when the beam reflects on the boundary.
\cite{ledizes2020} showed that a viscous correction with an amplitude $O(E^{1/6})$ smaller than the main beam is created at reflection.
He further showed that it takes the  form of equation \eqref{eq: self-similar-expression}  with a larger index, $m+1$.
In the present closed geometry, these corrected waves propagate on the closed circuit and, upon
summation, yield an expression of the form \eqref{eq: final-linear-solution} with an amplitude of order $E^{1/4}$ and functions $G_{m+1}$ replacing $G_m$.
Corrections of this form are expected to be created at the points $P_3$ and $P_7$ for both northward and southward beams.

For the reflections at $P_2$ and $P_6$, an even larger viscous correction is expected.
\citet{ledizes2024} indeed showed that a corrected beam of amplitude $O(E^{1/6})$ (that is $O(E^{1/12})$ smaller the main beam), was created at the critical point.
This beam has also the similarity form \eqref{eq: self-similar-expression} but an index $m+5/4$. As for the corrections generated at $P_3$ and $P_7$,  this correction is expected to accumulate as it propagates
along the closed circuit, resulting in an expression  of the form \eqref{eq: final-linear-solution} with an amplitude of order $E^{1/6}$ and functions $G_{m+5/4}$ instead of $G_m$.
Naturally, these first order corrections will themselves generate higher corrections of order $E^{1/4}$ as they reflect at $P_2$ and $P_6$, and of order $E^{1/3}$ as they reflect
at $P_3$ and $P_7$.
As a result, we expect the higher-order corrections to the harmonic solution \eqref{eq: final-linear-solution} to be composed, up to $O(E^{5/12})$, of sum of functions  $G_{m'}^N$ and $G_{m'}^S$  with different values $m'$.
This has an important consequence:  the harmonic solution including these viscous corrections  satisfies
equations \eqref{eq: v1phi2} and \eqref{eq: v1r} up to $O(E^{5/12})$.

\section{Mean flow corrections}\label{sec: results}

Our investigation of mean flow corrections proceeds systematically through two main aspects: the global response driven by oscillating boundary layers, and the localized interactions at beam crossing regions. For each interaction region, we present detailed comparisons between numerical and asymptotic solutions, accompanied by rigorous analysis of their scaling behaviors with respect to the Ekman number. This analysis encompasses the scaling of beam characteristics and velocity amplitudes.

\begin{figure}
	\centering
	\includegraphics[width=0.7\textwidth]{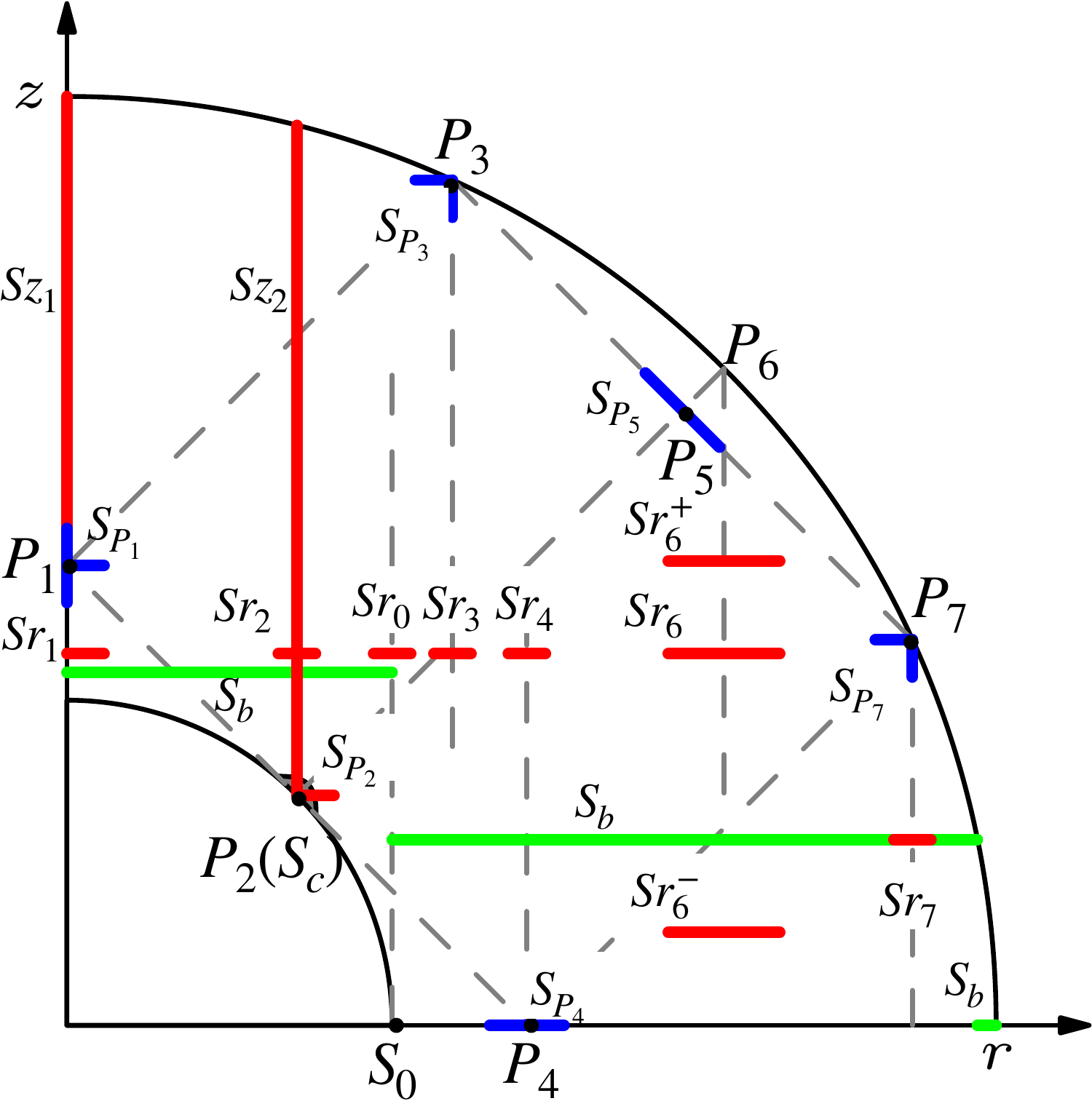}
	\caption{Schematic diagram of the positions of the slices used for comparing theoretical and numerical solutions (blue cuts). Green cuts are used to show the global bulk response. Red slices correspond to cases where only the numerical solution is discussed.}
	\label{fig: cuts-pos}
\end{figure}

To facilitate a clear visualization of the analysis locations, we employ the color-coded cutting system shown in figure~\ref{fig: cuts-pos}. Blue markers indicate the locations where theoretical solutions are obtained and compared to numerical solutions, red markers designate the positions where only numerical results are analysed, and green markers highlight the cuts along which bulk flow solutions are compared.

\subsection{Solution in the bulk generated from the boundary layers}\label{sec: bulk}

In this section, we consider the mean flow correction driven by oscillating boundary layers on the inner core.
Such a correction was already examined by \citet{sauret2013} in the absence of critical latitudes and inertial waves.
They determined that the mean zonal flow within a spherical shell could be directly inferred from the analytical solutions derived for a full sphere.
They showed that an azimuthal velocity of order $E^{0}$ was generated in the bulk, $v_{0\phi}=\mathcal{F}(\rho;\omega)$ in their equation (4.27).
Furthermore, they extended their analysis to cases where $0<\omega<2$, a condition in which inertial waves emerge and have a weak nonlinear effect on the fluid.
Building upon this foundational work, \citet{cebron2021} expanded the theory to include non-homoeoidal spheroidal shells, accounting for variations in the amplitudes of libration forcing at different boundaries.
They succinctly expressed the rotation velocity of the mean zonal flow in dimensionless form, presented as equation (5.5) in their publication.

\begin{figure}
	\centering
	\includegraphics[width=\textwidth]{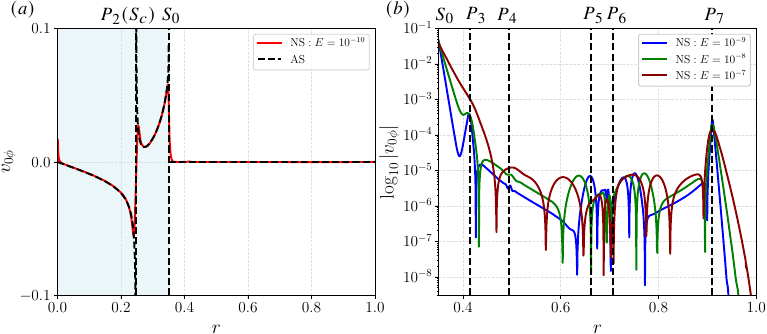}
	\caption{$(a)$ Mean azimuthal velocity $v_{0\phi}$ in the bulk as predicted by numerical computations (red solid line, $E = 10^{-10}$) and theoretical results (black dashed line, based on \citet{sauret2013} and \citet{cebron2021}). The light blue background highlights the interior region inside the tangent cylinder, located at $r = 0.35$. $(b)$ Logarithm mean azimuthal velocity outside the tangent cylinder from numerical computations at different Ekman numbers. Inside the tangent cylinder, cuts are taken at $S_{b}$ and $z = 0.4$. For the outer region, cuts are taken at $z = 0.2$ initially, and then on the equator once the cut reaches the outer boundary, see Figure~\ref{fig: cuts-pos}. The directory containing the data and the Jupyter notebook used to compute the asymptotic theory and generate this figure can be accessed at \url{https://www.cambridge.org/S0022112025109841/JFM-Notebooks/files/Figure5/Figure5.ipynb}.}
	\label{fig: bulk-compare}
\end{figure}

In our study, we use equation~(5.5) from \citet{cebron2021}, illustrating the theoretical predictions with a black dashed line in panel (a) of figure~\ref{fig: bulk-compare}.
Our numerical computations at $E=10^{-10}$ show excellent agreement with this theoretical prediction.
This confirms that nonlinear effects within the Ekman boundary layer remain one of the important sources of mean zonal flow within the shell.
Additionally, we identify two critical lines: $r=\eta/\sqrt{2}$, which marks the critical latitude, and $r=\eta$, denoting the cylinder tangent to the inner core boundary.
Our numerical computations confirm that equation~(5.5) from \citet{cebron2021} remains applicable even in the presence of inertial waves.

\begin{figure}
	\centering
	{\includegraphics[width=\textwidth]{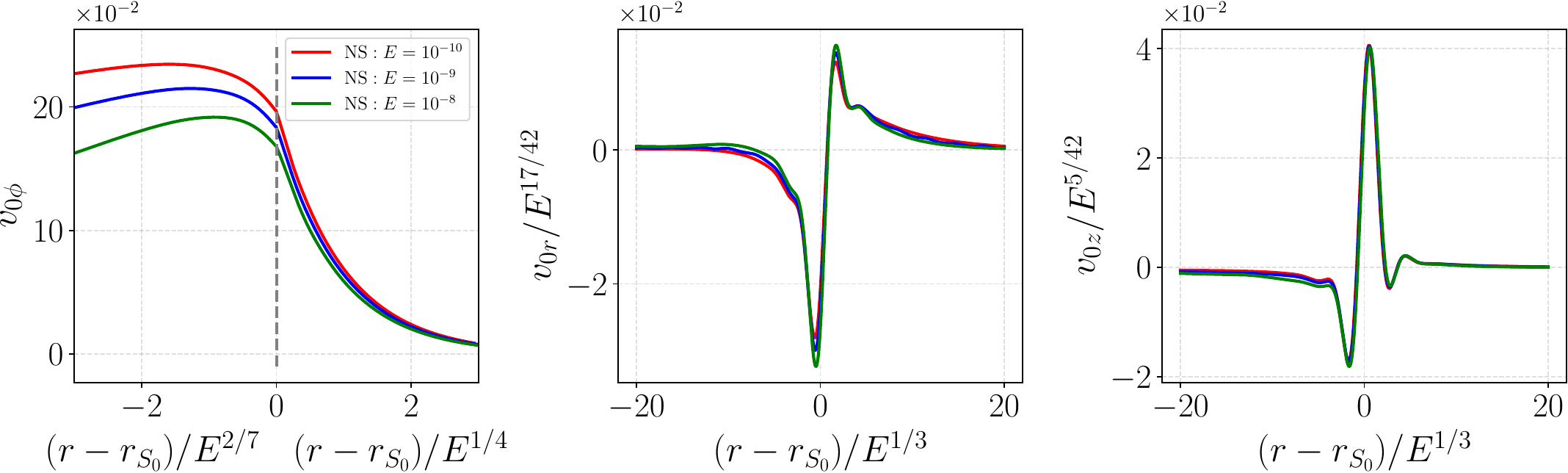}}
	\caption{The length and amplitude scaling at $Sr_{0}$ are revealed by the velocity profiles of three Ekman numbers. The different scalings observed in the azimuthal velocity component $v_{0\phi}$ originate from the Stewartson layer.}
	\label{fig: num-p3}
\end{figure}

In the band issued from $S_0$, which is tangent to the inner core, we expect Stewartson layers  characterized by three distinct widths:  $E^{2/7}$ for  the internal layers, $E^{1/3}$ for the inner layer,  and $E^{1/4}$
for the external layers \citep{stewartson1966,dormy2007,sauret2013}.
In figure~\ref{fig: num-p3}(a), the azimuthal velocity  is plotted  using the internal and external layer variables. The external scaling is perfectly recovered for this component.
For the radial and axial velocity components,  the scalings in $E^{17/42}$ and $E^{5/42}$, that are predicted  by the theory \citep{sauret2013}  in the inner layer, are compatible with the numerical results,
as shown in figures~\ref{fig: num-p3}(b,c).

It is important to note that previous theoretical frameworks generally assumed that the velocity outside the tangent cylinder remains zero when forcing is applied at the inner core boundary only.
Contrary to this assumption, our numerical results shown in figure~\ref{fig: bulk-compare}(b) reveal that this is not the case when inertial waves are present.
We indeed observe a bulk azimuthal flow of order $E^{1/3}$. Also visible are distinct peaks in the velocity profiles within the outer shell. These peaks correspond to specific regions where shear layers emitted from the critical latitude interact, either due to reflection on boundaries or through intersection within the volume. Although weak, these bands appear to exhibit the same scaling in  $E^0$ as the mean flow generated by the boundary layer on the inner core.
This observation provides important context for the subsequent research questions: What is the origin of these weak nonlinear zonal beams? How can we determine the corresponding amplitude and beam width? Addressing these questions constitutes one of the objectives of the present study.
In the following sections, we explain the origin of these weak nonlinear zonal beams and demonstrate the observed Ekman scaling.

\subsection{Solution generated from the interaction along the beam}
\label{sec: beam}

The equations \eqref{eq: nonlinear-governing}  governing the mean flow correction read in cylindrical coordinates
\begin{subequations}
	\begin{align}
		-2v_{0\phi}+\frac{ \partial p_{0} }{ \partial r }                                            & =-\mathcal{N}_{0r} + E\left(\Delta -\frac{1}{r^2}\right)v_{0r},      \\
		2v_{0r}                                                                                      & =-\mathcal{N}_{0\phi} +E\left(\Delta -\frac{1}{r^2}\right)v_{0\phi}, \\
		\frac{ \partial p_{0} }{ \partial z }                                                        & =-\mathcal{N}_{0z} +E\Delta v_{0z} ,                                 \\
		\frac{1}{r}\frac{ \partial  }{ \partial r } (rv_{0r})+\frac{ \partial v_{0z} }{ \partial z } & =0 ,
	\end{align}\label{eq: mean-flow}
\end{subequations}
where the Reynolds stress $\mathcal{N}_0=(\mathcal{N}_{0r}, \mathcal{N}_{0\phi}, \mathcal{N}_{0z})$ is given by
\begin{subequations}
	\begin{align}
		\mathcal{N}_{0r}    & = v_{1r}\frac{ \partial v_{1r}^{*} }{ \partial r } +v_{1z}\frac{ \partial v_{1r}^{*} }{ \partial z } - \frac{v_{1\phi}v_{1\phi}^{*}}{r}+\mathrm{c}.\mathrm{c} ~  ,  \\
		\mathcal{N}_{0\phi} & = v_{1r}\frac{ \partial v_{1\phi}^{*} }{ \partial r } +v_{1z}\frac{ \partial v_{1\phi}^{*} }{ \partial z } + \frac{v_{1r}v_{1\phi}^{*}}{r}+\mathrm{c}.\mathrm{c} ~, \\
		\mathcal{N}_{0z}    & =v_{1r}\frac{ \partial v_{1z}^{*} }{ \partial r } +v_{1z}\frac{ \partial v_{1z}^{*} }{ \partial z } +\mathrm{c}.\mathrm{c} ~ .
	\end{align}\label{eq: rss-cyl}
\end{subequations}

The Reynolds stress expression can be simplified along the beam using the properties \eqref{eq: v1phi2} and \eqref{eq: v1r} that the harmonic solution satisfies.
It gives using \eqref{eq: v1phi2} and \eqref{eq: v1r}
\bsea
&\mathcal{N}_{0r}    & \sim  \frac{\partial }{\partial r}\left( | v_{1r}|^2 - | v_{1z}| ^2\right) - 4   \frac{| v_{1r} |^2 }{r}  ~  , \\
&	\mathcal{N}_{0\phi} & \sim  2   \frac{\partial }{\partial z} \operatorname{Im}(v_{1r} v_{1z}^*)  ~,\\
&	\mathcal{N}_{0z}    &  \sim  -\frac{\partial }{\partial z}\left( | v_{1r}|^2 - | v_{1z}| ^2\right)  ~ .
\esea

If we neglect the viscous terms, a particular solution to equations \eqref{eq: mean-flow} can  be obtained as
\bsea
&v_{0r} &=  -  \frac{\partial }{\partial z} \operatorname{Im}(v_{1r} v_{1z}^*) ,\\
&v_{0\phi} &=   \frac{\partial }{\partial r}\left( | v_{1r}|^2 - | v_{1z}| ^2\right) - 2   \frac{| v_{1r} |^2 }{r} , \\
&v_{0z} & =     \frac{1}{r}\frac{\partial }{\partial r} r \operatorname{Im}(v_{1r} v_{1z}^*) .
\label{exp:v0gen}
\esea

As the harmonic velocity ${\bf v}_1$ varies with respect to a spatial variable that scales as $E^{1/3}$
for the main beam, and as $E^{1/6}$ for the secondary beam, the viscous terms associated with
the solution (\ref{exp:v0gen}) remain small. This expression is therefore expected to be valid
along the beam.

However, this solution is small, of order $E^{1/6}$, when evaluated using
\eqref{eq: final-linear-solution}, and of order $E^{1/2}$, when evaluated using \eqref{eq: linear-weak}.
It becomes  large only when two beams  intersect, that is in the local regions $P_\beta$. In the next subsections,
we provide the scaling and the expression of the solution in the three
typical interaction regions.

\subsubsection{Solution  in the local regions $P_3$, $P_4$ and $P_7$}
\label{sec: P347}

Close to those points, the particular solution reduces,  at leading order,  to
\begin{equation}
	\mathbf{v}_0  \sim \frac{1}{r}E^{-1/6} \mathbf{\tilde{v}}_0 \ ,
\end{equation}
with
\begin{equation}
	\begin{pmatrix}
		\tilde{v}_{0r}    \\[10pt]
		\tilde{v}_{0\phi} \\[10pt]
		\tilde{v}_{0z}
	\end{pmatrix}=\begin{pmatrix}
		\displaystyle{-\frac{1}{\sqrt{2}}  \frac{ \partial \widetilde{\mathcal{N}}_{P_\beta} }{ \partial \tilde{z}_\beta }      } \\[8pt]
		\displaystyle{  \frac{ \partial  \widetilde{\mathcal{M}}_{P_\beta} }{ \partial \tilde{r}_\beta }  }                       \\[8pt]
		\displaystyle{\frac{1}{\sqrt{2}}   \frac{ \partial  \widetilde{\mathcal{N}}_{P_\beta} }{ \partial \tilde{r}_\beta }      }
	\end{pmatrix} .
	\label{eq: veloP}
\end{equation}
where the functions $\widetilde{\mathcal{M}}_{P_{\beta}}$ and  $\widetilde{\mathcal{N}}_{P_{\beta}}$ are given by
\begin{subeqnarray}
	\widetilde{\mathcal{M}}_{P_{3}} =  2\mathrm{Re}\widetilde{\mathcal{Q}}_3~,~~  & \widetilde{\mathcal{M}}_{P_{4}} = 2\mathrm{Re}\widetilde{\mathcal{Q}}_4 ~,~~& \widetilde{\mathcal{M}}_{P_{7}} = - 2\mathrm{Re}\widetilde{\mathcal{Q}}_7~, \\
	\widetilde{\mathcal{N}}_{P_{3}} = -2\mathrm{Im}\widetilde{\mathcal{Q}}_3 ~,~~ & \widetilde{\mathcal{N}}_{P_{4}} = -2\mathrm{Im}\widetilde{\mathcal{Q}}_4 ~,~~& \widetilde{\mathcal{N}}_{P_{7}} =  -2\mathrm{Im}\widetilde{\mathcal{Q}}_7~,
	\label{exp: Np}
\end{subeqnarray}
with
\be
\widetilde{\mathcal{Q}}_\beta(\tilde{r}_\beta,\tilde{z}_\beta) =(\tilde{B}_{\beta}^{Ni}- \tilde{B}_{\beta}^{Sr} )( \tilde{B}_{\beta}^{Nr*} - \tilde{B}_{\beta}^{Si*}) \ .
\label{eq: NB}
\ee
As it can be seen on these expressions, only four nonlinear contributions remain.
They are all confined to the local interaction regions.
This result aligns with the outcomes of our numerical computations and meets our expectations.
This further supports the idea that the main sources of zonal flow are constrained to the $O(E^{1/3})$   regions where incident and reflected beams interact.

From \eqref{eq: veloP}, one can also obtain the normal and tangential velocity associated with this solution:
\begin{equation}
	\tilde{v}_{0n}=-\frac{1}{\sqrt{2}}   \frac{ \partial  \widetilde{\mathcal{N}}_{P_\beta} }{ \partial \tilde{x}_{t} } ,\quad \tilde{v}_{0t}=\frac{1}{\sqrt{2}}   \frac{ \partial  \widetilde{\mathcal{N}}_{P_\beta}  }{ \partial \tilde{x}_{n} }  \ .
\end{equation}
Thanks to \eqref{eq:Bxn=0}, one can see  that expression \eqref{exp: Np} with \eqref{eq: NB} reduces at the boundary ($\tilde{x}_n=0$)  to $ \widetilde{\mathcal{N}}_{P_\beta} (\tilde{x}_n=0) = 0$. This implies that
the normal velocity vanishes at the boundary.  The  non-penetration condition is then automatically satisfied. This has an important consequence. It means that the particular solution \eqref{eq: veloP} is the leading-order solution in each local region around $P_\beta$.
It is this expression that will be compared to the numerical solution close to each $P_\beta$.
Since the functions $ \widetilde{\mathcal{M}}_{P_\beta}$ and  $\widetilde{\mathcal{N}}_{P_\beta}$ are both localized, all the velocity components are also localized near $P_\beta$ at this order.

\begin{figure}
	\centering
	\includegraphics[width=\textwidth]{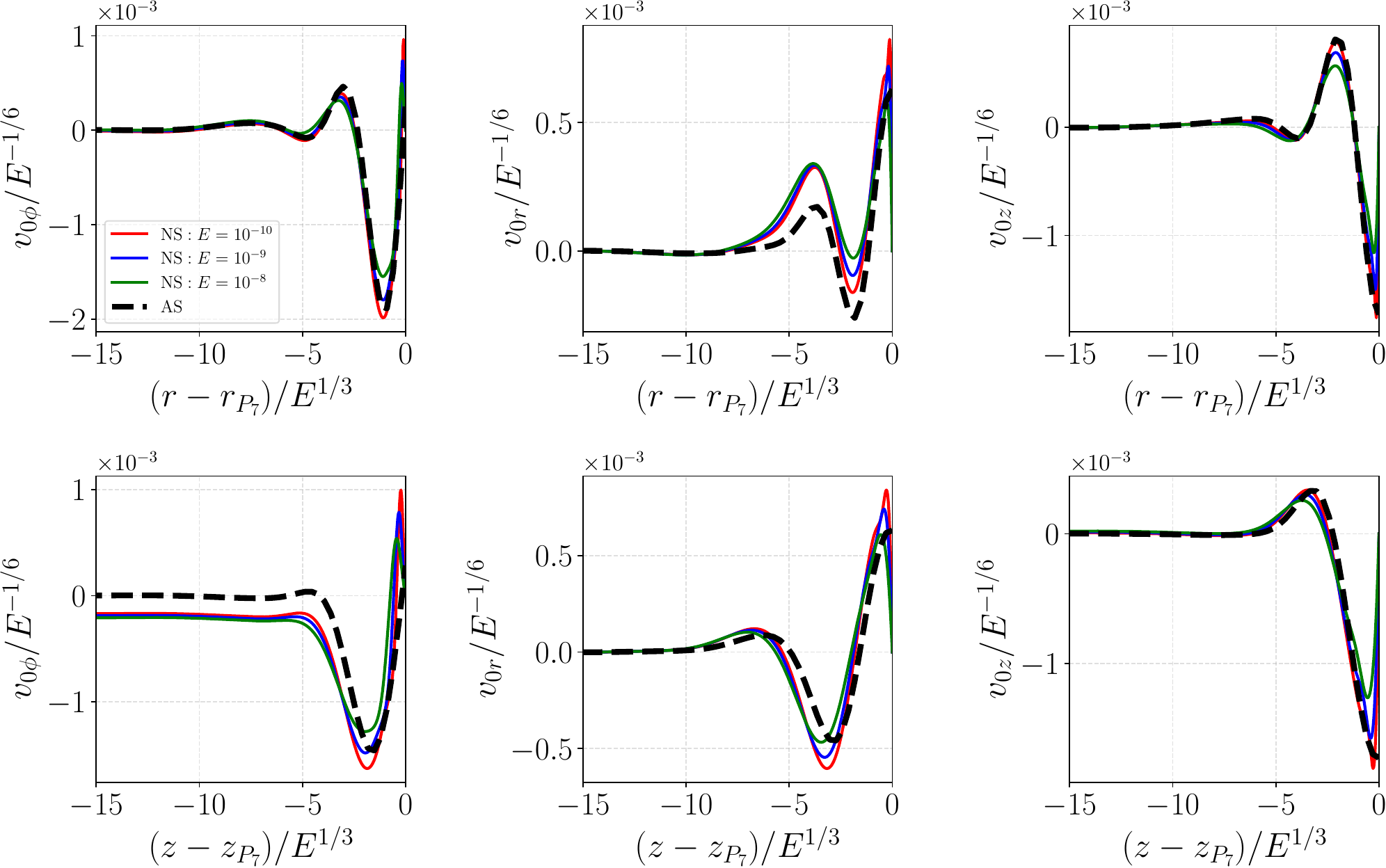}
	\caption{Three velocity components profiles of the asymptotic solution ($AS$) and the numerical solution for three different Ekman numbers at local region $P_7$ with two direction cuts ($S_{P_7}$ in figure~\ref{fig: cuts-pos}). The directory containing the data and the Jupyter notebook used to compute the asymptotic theory and generate this figure can be accessed at \url{https://www.cambridge.org/S0022112025109841/JFM-Notebooks/files/Figure7/Figure7.ipynb}.}
	\label{fig: asy-num-p7}
\end{figure}

Figure~\ref{fig: asy-num-p7} compares the profiles of three asymptotic velocity components with numerical solutions for three different Ekman numbers near $P_7$.
The comparison is performed on the vertical and horizontal slices $S_{P_7}$, as depicted in figure~\ref{fig: cuts-pos}.
As expected, the asymptotic theoretical results increasingly align with the numerical data as the Ekman number decreases.
However, discrepancies are observed in the radial velocity, $v_{0r}$, and the axial velocity, $v_{0z}$, near the boundary.
These discrepancies arise because the numerical boundary condition employs a no-slip condition, causing the numerical solutions to approach zero at the boundary, while only the non-penetration condition is satisfied by
the asymptotic solution.
As demonstrated in \citet{ledizes2020}, another approximation can be constructed in a viscous boundary layer of $O(E^{1/2})$ width to capture this behaviour.

In addition, the azimuthal velocity along the $z$-axis, parallel to the rotation axis, tends to zero in the asymptotic theory as one moves away from the local interaction point.
However, numerical results display a small amplitude zonal flow in the bulk (see the bottom left panel of figure~\ref{fig: asy-num-p7}).
This is also evident in figure~\ref{fig: contourmap}, where several weak  bands are visible throughout the bulk.
The scaling of these bands will be discussed in the next section, using the results given in appendix~\ref{app:band}.

\begin{figure}
	\centering
	\includegraphics[width=\textwidth]{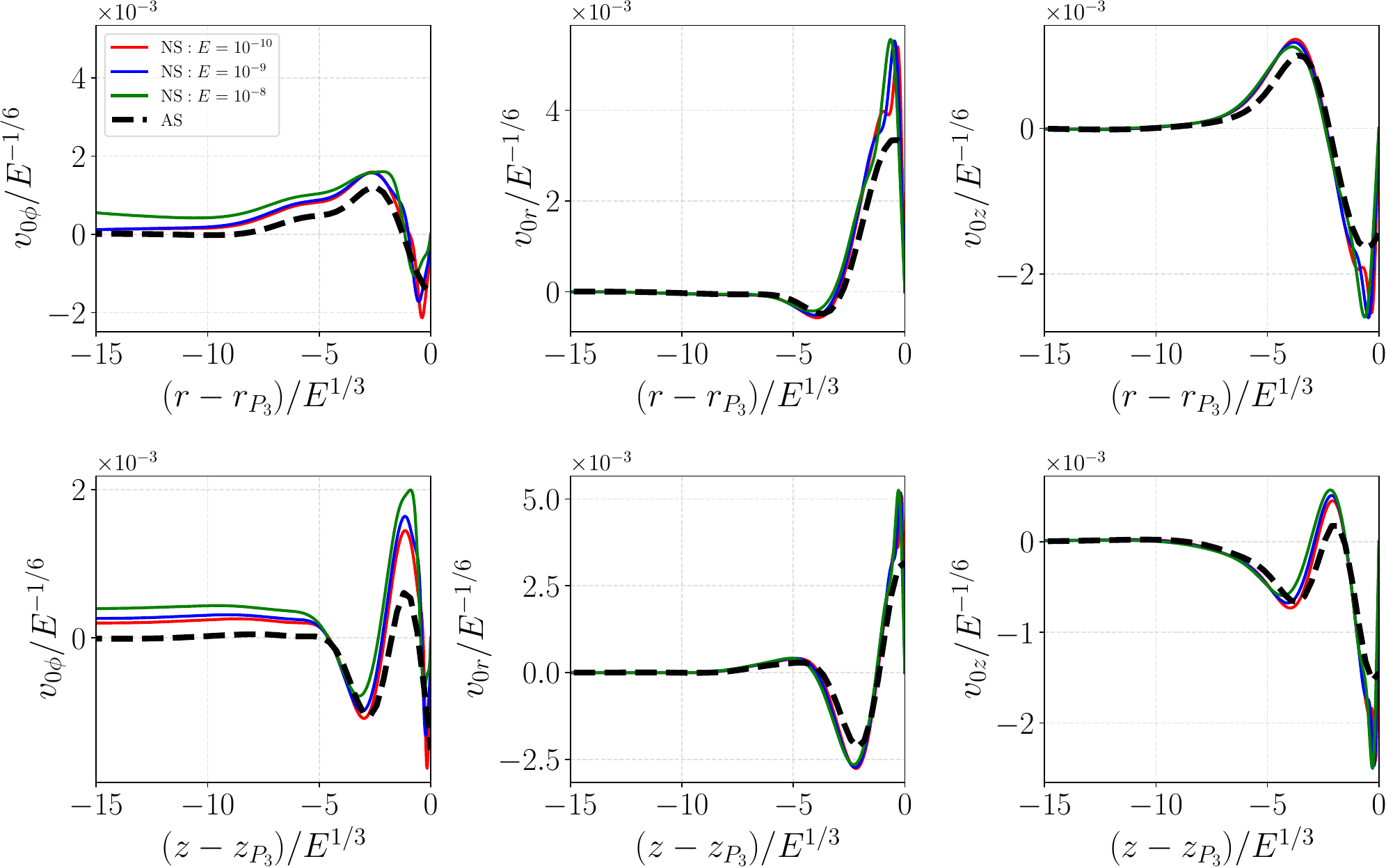}
	\caption{Same caption as figure \ref{fig: asy-num-p7} but at local region $P_3$ (Cuts at $S_{P_3}$). The directory containing the data and the Jupyter notebook used to compute the asymptotic theory and generate this figure can be accessed at \url{https://www.cambridge.org/S0022112025109841/JFM-Notebooks/files/Figure8/Figure8.ipynb}.}
	\label{fig: asy-num-p4}
\end{figure}

\begin{figure}
	\centering
	\includegraphics[width=0.8\textwidth]{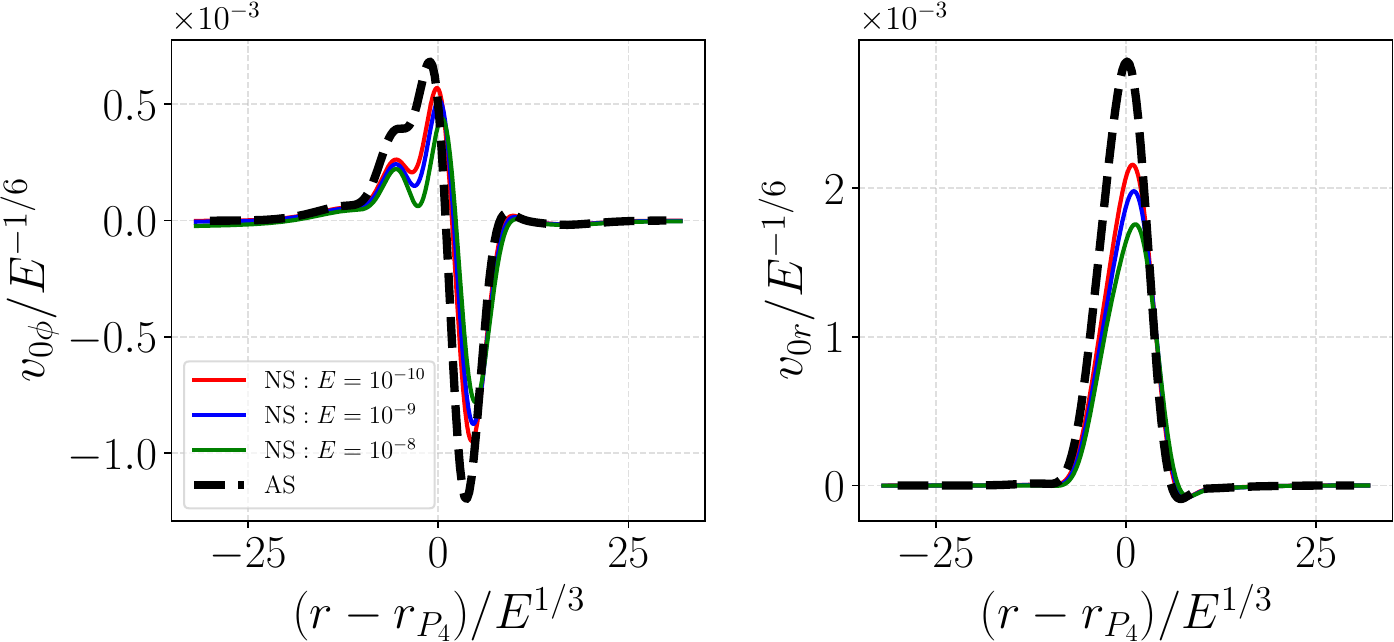}
	\caption{Comparison of asymptotic solutions and numerical results of two velocity components ($v_{0z}=0$) at local region $P_4$. (Cuts position at $S_{P_4}$). The directory containing the data and the Jupyter notebook used to compute the asymptotic theory and generate this figure can be accessed at \url{https://www.cambridge.org/S0022112025109841/JFM-Notebooks/ files/Figure9/Figure9.ipynb}.}
	\label{fig: asy-num-p5}
\end{figure}

Comparisons between two similar interaction regions, $P_3$ and $P_4$, and the numerical results are presented in figures~\ref{fig: asy-num-p4} and~\ref{fig: asy-num-p5}, respectively.
The observations from figure~\ref{fig: asy-num-p7} regarding differences in the boundary layer and the azimuthal velocity in the bulk are also applicable here.
Notably, near $P_4$, located on the equatorial plane, the axial velocity $v_{0z}$ remains zero by symmetry.
Additionally, improvements in the comparison with decreasing Ekman numbers are observed, consistent with the findings at other local positions.

At the three specific locations, $P_7$, $P_3$, and $P_4$ the asymptotic analysis yielded consistent scaling behaviours: the interaction region scales as $E^{1/3}$, and the amplitude scales as $E^{-1/6}$ for all three velocity components.
Note however that the rescaled velocity remains small, of order $10^{-3}$, at each of these points. This explains why these regions are not more visible on the contour map shown in figure~\ref{fig: contourmap}.

\subsubsection{Solution in the local regions $P_5$ and $P_6$}
\label{sec: P5}

The secondary weak beams encounter the main critical beam in the bulk at $P_5$. This  case is  interesting because the scalings of the two beams differ.

The mean flow correction is weaker and given by
\begin{equation}
	\mathbf{v}_{0}\sim\frac{1}{r}E^{0}\mathbf{\breve{v}}_{0} \ ,
\end{equation}
with
\begin{equation}
	\begin{pmatrix}
		\breve{v}_{0r}    \\[10pt]
		\breve{v}_{0\phi} \\[10pt]
		\breve{v}_{0z}
	\end{pmatrix}
	=\begin{pmatrix}
		\displaystyle{-\frac{1}{\sqrt{2}}  \frac{ \partial \breve{\mathcal{N}}_{P_5} }{ \partial \tilde{z}_5 }      } \\[8pt]
		\displaystyle{  \frac{ \partial  \breve{\mathcal{M}}_{P_5} }{ \partial \tilde{r}_5 }  }                       \\[8pt]
		\displaystyle{\frac{1}{\sqrt{2}}   \frac{ \partial  \breve{\mathcal{N}}_{P_5} }{ \partial \tilde{r}_5 }      }
	\end{pmatrix}, \label{eq: nonlinear-p5}
\end{equation}
where
\begin{align}
	\breve{\mathcal{M}}_{P_5} & =	2\operatorname{Re} \left(  ( \breve{B}^{N}- \breve{B}^{S}) ( \tilde{B}_{5}^{N*} -\tilde{B}_{5}^{S*}) \right) , \\
	\breve{\mathcal{N}}_{P_5} & =	 -2 \operatorname{Im}\left( ( \breve{B}^{N}- \breve{B}^{S}) ( \tilde{B}_{5}^{N*} -\tilde{B}_{5}^{S*}) \right).
\end{align}
These expressions are similar to the one obtained near point $P_4$, where the $Nr$ and $Si$ beams at $P_4$ are now the strong $S$ and $N$ beams propagating between $P_3$ and $P_7$, while $Ni$ and $Sr$ beams at $P_4$ are the weak $S$ and $N$ beams propagating between $P_2$ and $P_6$ (see figure~\ref{fig: configure}(b)).
Note that because $\breve{B}^S$ and $\breve{B}^N$ are functions of the slow variable $\breve{x}_\perp = E^{1/6} (\tilde{z}_5-\tilde{r}_5)/\sqrt{2}$, the spatial derivatives in \eqref{eq: nonlinear-p5} should be applied to
$\tilde{B}_{5}^{S}$ and $\tilde{B}_{5}^{N}$ only.

\begin{figure}
	\centering
	\includegraphics[width=\textwidth]{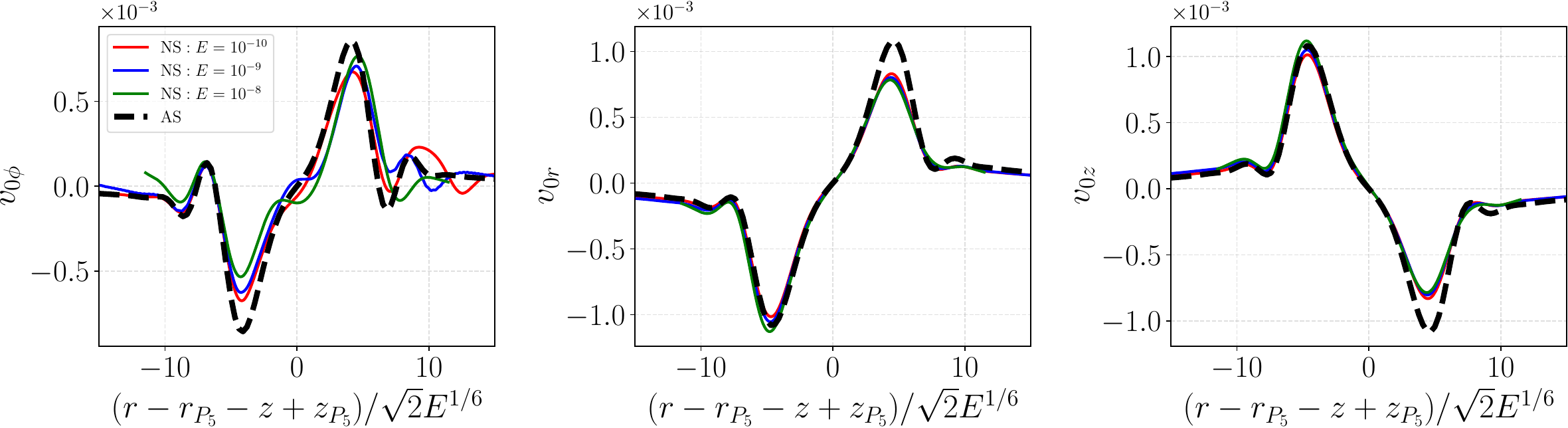}
	\caption{Comparison of the asymptotic solutions and numerical results of velocity profiles at local region $P_{5}$ (Cuts at $S_{P_5}$). The directory containing the data and the Jupyter notebook used to compute the asymptotic theory and generate this figure can be accessed at \url{https://www.cambridge.org/S0022112025109841/JFM-Notebooks/files/Figure10/ Figure10.ipynb}.}\label{fig: p5}
\end{figure}

In figure~\ref{fig: p5}, we compare the asymptotic solution from equation~\eqref{eq: nonlinear-p5} with numerical results for three different Ekman numbers. The secondary weak beam, characterized by its larger width scaling in $E^{1/6}$, is analyzed by taking profile cuts parallel to the critical beam from $P_3$ to $P_7$ (see figure~\ref{fig: cuts-pos}, $S_{P_5}$). Due to the anti-symmetry of the weak beam, the cuts in the perpendicular direction
(that is on the $z = r$ line), are zero.

The width of the local region scales as $E^{1/6}$, while the other direction scales as $E^{1/3}$. The amplitude scaling is $E^{0}$, which is significantly weaker than in other local interaction regions. Despite the two beams having different scaling factors, the asymptotic method demonstrates overall accuracy. The asymptotic theory matches the numerical solutions well, exhibiting convergence as the Ekman number decreases.

However, some perturbations are observed on the right side of figure~\ref{fig: p5}, specifically in the $v_{0\phi}$ component. These perturbations arise because the cut profiles are influenced by the weak band issued from the point $P_6$,  as shown in figure~\ref{fig: bulk-compare}(b).

In the local region $P_6$, a particular solution of the same form as \eqref{eq: nonlinear-p5} can be obtained. 
However, as we will see in \S \ref{sec: num-scalings} and Appendix~\ref{app:band}, this particular solution is modified by an additional homogeneous solution,
which describes the mean flow correction associated with the vertical band originating from $P_6$. Therefore, it cannot be directly compared to numerical results.

\subsubsection{Solution in the local region $P_1$}\label{sec: P1}

\begin{figure}
	\centering
	\includegraphics[width=0.66\textwidth]{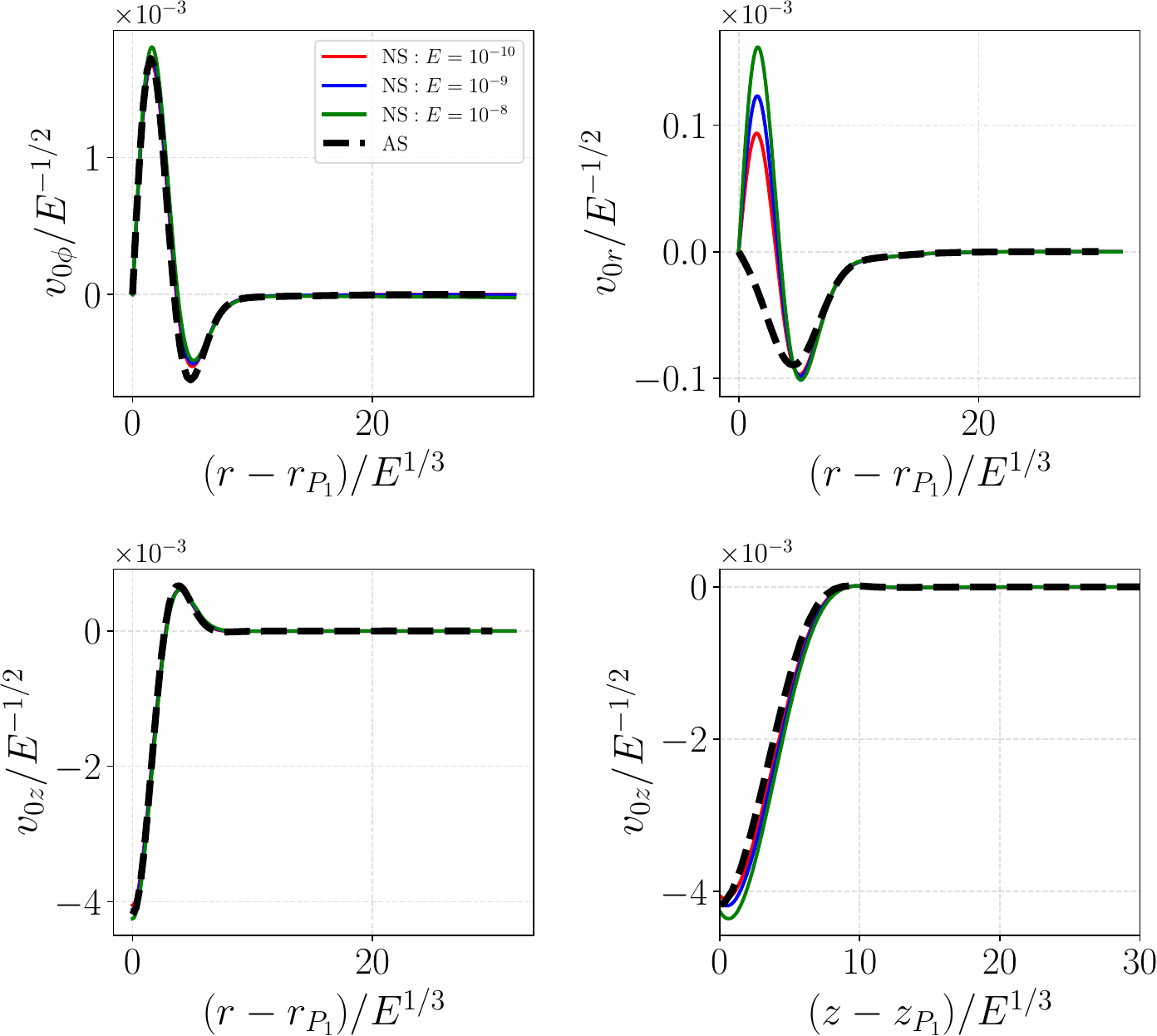}
	\caption{Comparison of the asymptotic solutions and numerical results of velocity profiles at local region $P_1$. (Cuts position at $S_{P_1}$). The directory containing the data and the Jupyter notebook used to compute the asymptotic theory and generate this figure can be accessed at \url{https://www.cambridge.org/S0022112025109841/JFM-Notebooks/files/ Figure11/Figure11.ipynb}.}
	\label{fig: asy-num-p1}
\end{figure}

For the point on the rotation axis, we must use the expressions \eqref{eq: p1-linear-velo} for the harmonic solution.
We obtain the following expression for the mean flow
\begin{equation}
	\mathbf{v}_0=E^{-1/2} \mathbf{\hat{v}}_{0} \ ,
\end{equation}
with
\begin{subequations}\label{eq: p1-velo}
	\begin{align}
		\hat{v}_{0r}
		 & = -\frac{1}{\sqrt{2}}\frac{\partial}{\partial \tilde{z}_1} \left\{\operatorname{Re}\left[ (Q^{N}_{1}+ Q^{S}_{1})(-Q^{N*}_{0}+Q^{S*}_{0}) \right]\right\}                                             \\[0em]
		\hat{v}_{0\phi}
		 & = \frac{1}{2}\frac{\partial}{\partial \tilde{r}_1} \left\{\left|Q^{N}_{1}+Q^{S}_{1}\right|^2 - \left|-Q^{N}_0 + Q^S_0\right|^2 \right\}   -  \frac{\left|Q^{N}_{1}+Q^{S}_{1}\right|^2 }{\tilde{r}_1}
		\\[0em]
		\hat{v}_{0z}
		 & = \frac{1}{\sqrt{2}}\left(\frac{\partial}{\partial \tilde{r}_1}+ \frac{1}{\tilde{r}_1} \right)\left\{\operatorname{Re}\left[ (Q^{N}_{1}+ Q^{S}_{1})(-Q^{N*}_{0}+Q^{S*}_{0}) \right]\right\}
		\ . \label{eq: nonlinear-p1-vphi}
	\end{align}
\end{subequations}
In figure~\ref{fig: asy-num-p1}, we compare the asymptotic solution from equation~\eqref{eq: p1-velo} with numerical results for three different Ekman numbers. The profiles analyzed are taken at $S_{P_1}$ and include one profile in the direction perpendicular to the rotation axis and another along the rotation axis (see figure~\ref{fig: cuts-pos}). Due to the axisymmetry of the system, only the axial velocity component is considered along the rotation axis, as the two  other components, $v_{0r}$ and $v_{0\phi}$, are identically zero.

The width of the local region scales with $E^{1/3}$, while the amplitude scales with $E^{-1/2}$. 
Although the overall agreement between the asymptotic solution and the numerical results is strong, particularly for the azimuthal component, a noticeable deviation is observed in the smaller $v_{0r}$ component near the axis. This discrepancy was already observed in the linear harmonic solution in \citet{he2022}. We suspect that it could come from higher-order corrections that could become non-negligible owing to the smallness of the solution at this leading order.

\section{Numerical scalings of the bands}
\label{sec: num-scalings}

In the previous section, we  analysed the mean flow corrections generated  in the interaction regions and in the bulk from the oscillating boundary layer on the inner core.
These represent   the dominant  contributions.
As illustrated in figure~\ref{fig: contourmap}, numerical results also reveal the presence of faint vertical beam-like bands within the bulk.
These bands originate from the local interaction regions. However, the leading-order mean flow corrections computed in these regions were shown to be localized.
The observed  bands must therefore be associated with higher-order effects.
Finding an approximation for the velocity field in the bands is thus also more complex, as it requires higher-order expansions of the harmonic solution.
Nevertheless, some general insight into the scaling and  structure of the velocity field in each band can  be obtained through asymptotic reasoning. These results are summarized in  Appendix~\ref{app:band} and
will be used to interpret the numerical results that are now presented.

\begin{figure}
	\centering
	{\includegraphics[width=\textwidth]{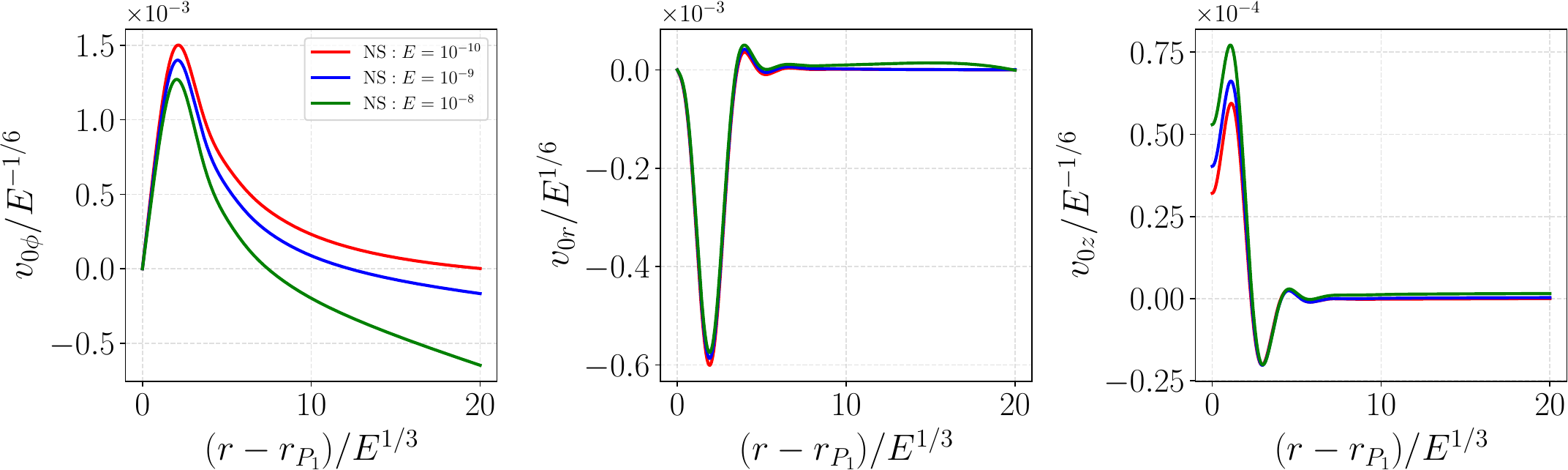}}
	\caption{Velocity scalings in the band originating from $P_1$. Variation along the cut $Sr_{1}$ (indicated in figure \ref{fig: cuts-pos}) for three different Ekman numbers $E=10^{-8}$, $E=10^{-9}$ and $E^{-10}$.
		(a) Azimuthal velocity. (b) Radial velocity. (c) Axial velocity. The scalings are those predicted by the theory (appendix \ref{app:band}).}
	\label{fig: num-p1}
\end{figure}

\begin{figure}
	\includegraphics[width=\textwidth]{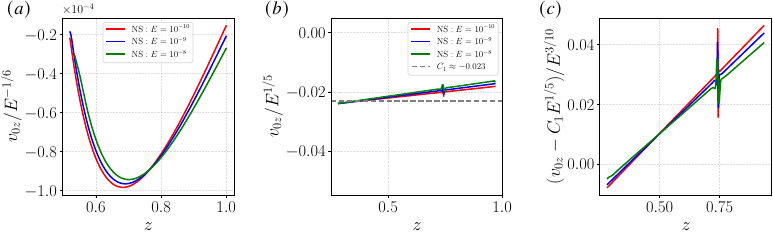}
	\caption{Vertical dependency of the bands originating from $P_1$ and $P_2$. (a): Band from $P_1$. Axial velocity along the axis (cut $Sz_1$), demonstrating a nonlinear dependency with respect to the vertical coordinate $z$; (b), (c):  Band from $P_2$.  Axial velocity versus $z$ along the cut $Sz_2$, showing a uniform behavior at the order $E^{1/5}$ (b) and a linear variation at the order $E^{3/10}$. The localized variation of the velocity close to $z\approx 0.75$ in (b) and (c) corresponds to the crossing with the main beam travelling between $P_1$ and $P_4$.}
	\label{fig: Vertical-p1-p2}
\end{figure}

Figure~\ref{fig: num-p1} displays the scaling of the vertical band aligned with the rotation axis, analysed along the cut $Sr_{1}$ shown in figure~\ref{fig: cuts-pos}.
Near $r=0$, the width of the band scales as $E^{1/3}$.
As explained in the appendix,  the amplitude of the axial and azimuthal velocity components  in the band should be $E^{1/3}$ smaller than in the local region around $P_1$.
This leads to expected scalings of $E^{-1/6}$ for $v_{0\phi}$ and $v_{0z}$, and $O(E^{1/6})$ for $v_{0r}$.
These scaling are confirmed by the numerical results shown in  figure \ref{fig: num-p1}.
The variation of the mean flow correction within this band along
the axial coordinate is also complex, as illustrated in figure~\ref{fig: Vertical-p1-p2}(a).

\begin{figure}
	\centering
	\includegraphics[width=\textwidth]{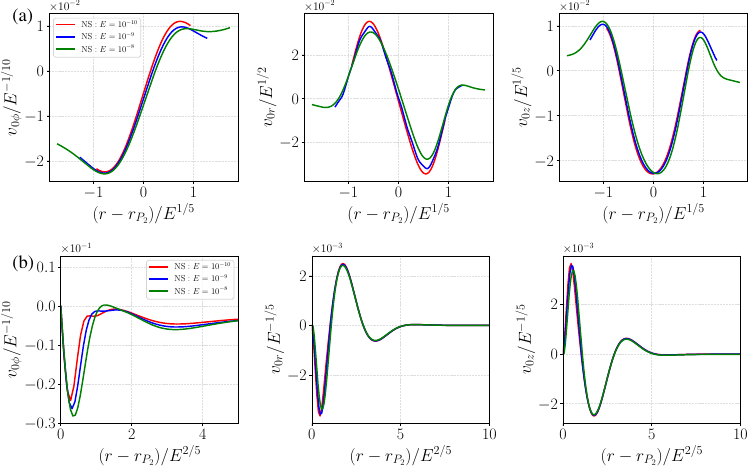}
	\caption{Velocity scalings in the local region $P_2$ (cut $S_{P_2}$ in (b)) and in the band originating from $P_2$ (cut $Sr_{2}$ in (a)).}
	\label{fig: num-p2}
\end{figure}
Figure~\ref{fig: num-p2} illustrates the scaling behaviours in two distinct regions: the critical latitude region (analysed at cut $S_{P_2}$) and the resulting zonal band in the bulk (analysed at cut $Sr_{2}$).
The horizontal width of the critical latitude region scales as  $E^{2/5}$, while the width of the associated band is $O(E^{1/5})$.
As explained in Appendix~\ref{app:band},  the scaling of the azimuthal velocity in the band depends on the Ekman pumping generated in the local region $P_2$, which is challenging to evaluate analytically.
However, the scalings of the different components are interrelated.  In figure~\ref{fig: num-p2}(a), we propose a scaling in $E^{-1/10}$ for $v_{0\phi}$, $E^{1/2}$ for $v_{0r}$ and $E^{1/5}$ for $v_{0z}$ which is consistent
with the expected structure of the mean flow correction in the band issued from $P_2$.
Figure \ref{fig: num-p2}(b) present scaling behavior in the local region $P_2$ that may be compatible with the observed scalings in the band.
In particular,  a scaling of $v_{0r}$ and $v_{0z}$ in $E^{-1/5}$ is expected to yield an Ekman pumping of order $E^{1/5}$, which matches the amplitude of the axial velocity observed in the band.
In figure~\ref{fig: Vertical-p1-p2}(b,c), we observe that the axial velocity within the band from $P_2$ is uniform along the axial direction
at  leading order $E^{1/5}$, but linear at the next order $E^{3/10}$, as predicted by the theory.

\begin{figure}
	\centering
	\includegraphics[width=\textwidth]{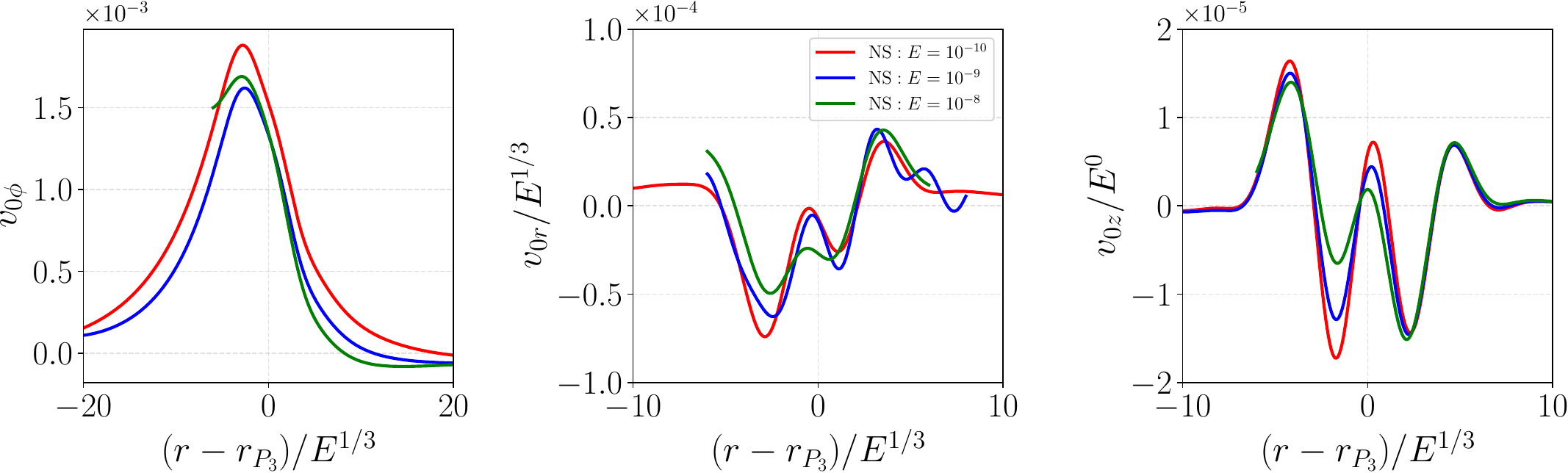}
	\caption{Same caption as figure \ref{fig: num-p1} but for the band originating from $P_3$ (cut $Sr_{3}$).}
	\label{fig: num-p4}
\end{figure}

\begin{figure}
	\centering
	\includegraphics[width=\textwidth]{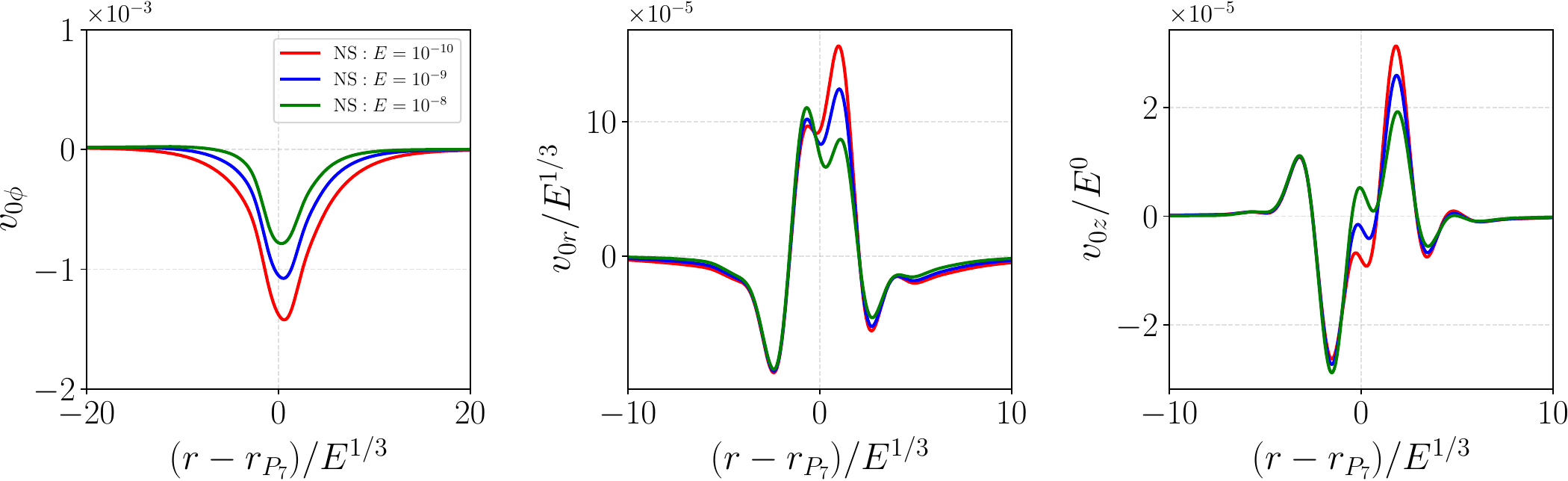}
	\caption{Same caption as figure \ref{fig: num-p1} but for the band originating from $P_7$ (cut $Sr_{7}$).}
	\label{fig: num-p7}
\end{figure}

\begin{figure}
	\centering
	\includegraphics[width=0.33\textwidth]{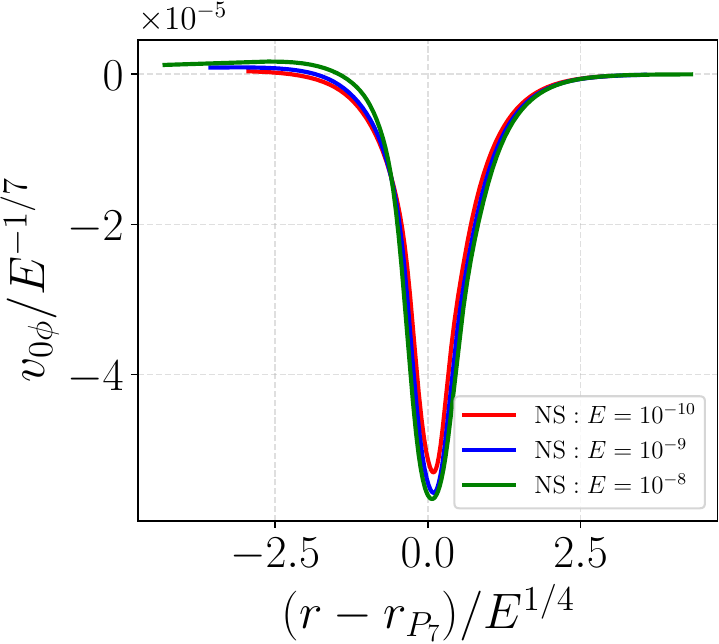}
	\caption{Alternative scaling for the azimuthal velocity in the band originating from $P_7$ (cut $Sr_{7}$).}
	\label{fig: num-p7_14}
\end{figure}

Figures~\ref{fig: num-p4} and~\ref{fig: num-p7} present the velocity profiles along cuts $Sr_{3}$ and $Sr_{7}$, respectively. We have applied the scalings predicted by the asymptotic analysis, namely, an amplitude
of order $E^0$ for the azimuthal and axial velocity components, and  $E^{1/3}$ for the radial velocity (see appendix \ref{app:band}).
These scalings are consistent with the numerical results. However,  for the azimuthal velocity in the band issued from $P_7$, an alternative scaling appears to better match the numerical data
(see figure~\ref{fig: num-p7_14}). At present, we have no theoretical justification for this alternative scaling.

\begin{figure}
	\centering
	\includegraphics[width=0.66\textwidth]{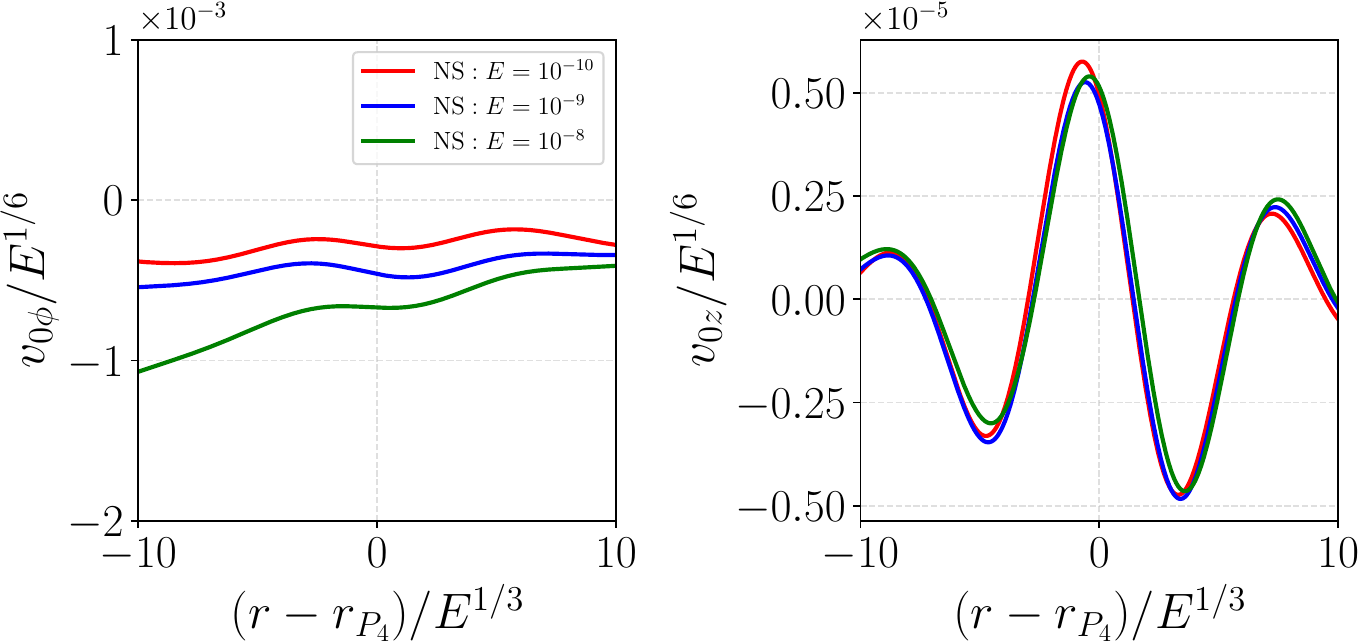}
	\caption{Same caption as figure \ref{fig: num-p1}, but for the band originating from $P_4$ (cut $Sr_{4}$). The very weak radial velocity that scales as $E^{1/2}$ has not been plotted.}
	\label{fig: num-p5}
\end{figure}

As discussed in Appendix \ref{app:band},  no significant band is expected to form from $P_4$, which was already evident in figure \ref{fig: contourmap}(e). In figure \ref{fig: num-p5}, we confirm that the numerical results align with the
theoretically predicted scaling of $E^{1/6}$ for both the azimuthal and axial velocity components.

\begin{figure}
	\centering
	\includegraphics[width=\textwidth]{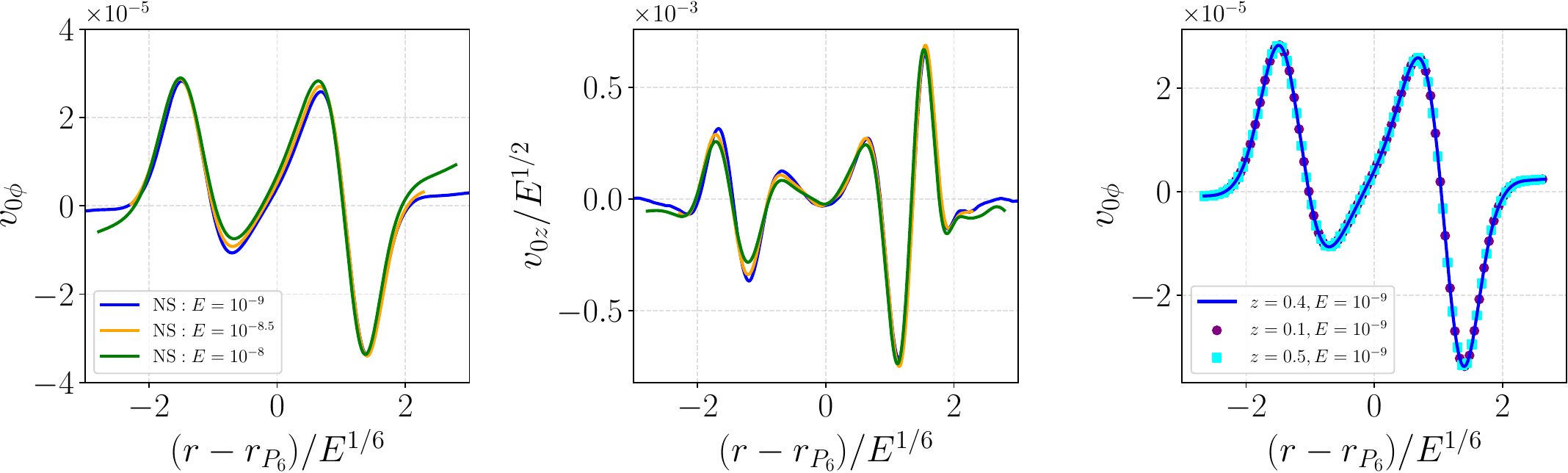}
	\caption{Velocity scalings in the band originating from $P_6$.
		(a) Azimuthal velocity versus $r$ along the cut $Sr_{6}$; (b) Axial velocity versus $r$ along the cut $Sr_{6}$; (c) Azimuthal velocity versus $r$ for different $z$, along the cut $Sr_{6}$, $Sr_{6}^{+}$ and $Sr_{6}^{-}$ at $z=0.4$, $z=0.5$ and $z=0.1$, respectively.}
	\label{fig: num-p62}
\end{figure}

A faint band originating from the critical point on the outer core is visible in figure  \ref{fig: contourmap}(a). This band is not associated with the local region $P_5$, which is not expected to generate any visible band, but rather with $P_6$.
As explained in Appendix~\ref{app:band}, the structure of the velocity field in this band is peculiar. Its width scales as $E^{1/6}$, matching with  the width of the local region $P_6$.  However,  unlike the
bands originating from $P_3$ and $P_7$, the axial velocity here is significantly weaker than the azimuthal velocity and varies linearly in $z$. The scaling used
in figure~\ref{fig: num-p62}(a,b) - $E^0$ for $v_{0\phi}$, $E^{1/2}$ for $v_{0z}$ - are those   predicted by the theoretical analysis. The radial velocity, which is expected to be $O(E^{2/3})$ is too weak to be correctly resolved, and has not been plotted.
In figure~\ref{fig: num-p62}(c), we demonstrate that the azimuthal velocity is uniform along the vertical direction, as expected from the theory. 

\section{Conclusion}\label{sec: conclusion}
\begin{figure}
	\centering
	\includegraphics[width=0.7\textwidth]{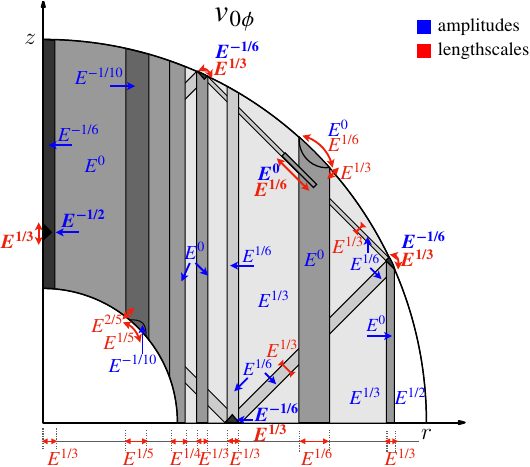}
	\caption{Ekman scaling of the  mean flow correction azimuthal velocity $v_{0\phi}$. The blue and red colors denote the amplitude and length scale scalings, respectively. The scalings indicated in bold have been validated by an asymptotic solution.}
	\label{fig: scaling-v0phi}
\end{figure}

\begin{figure}
	\centering
	\includegraphics[width=0.7\textwidth]{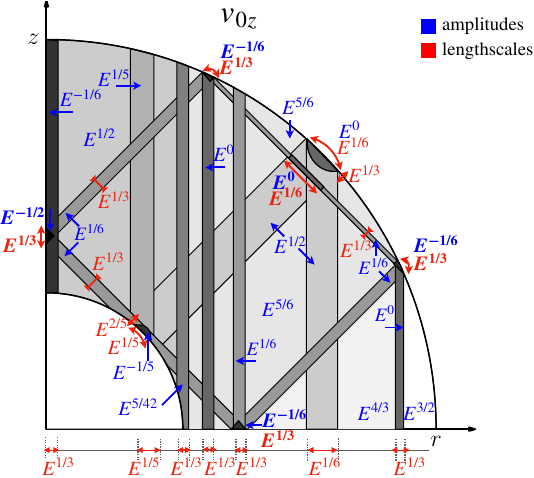}
	\caption{Same caption as in figure~\ref{fig: scaling-v0phi}, but shown for the velocity component $v_{0z}$.}
	\label{fig: scaling-v0z}
\end{figure}

\begin{figure}
	\centering
	\includegraphics[width=0.7\textwidth]{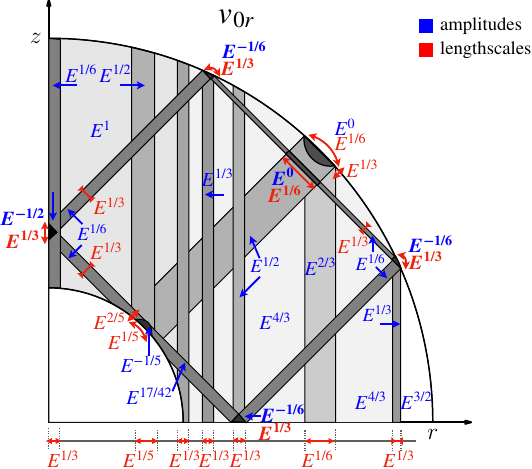}
	\caption{Same caption as in figure~\ref{fig: scaling-v0phi}, but shown for the velocity component $v_{0r}$.}
	\label{fig: scaling-v0r}
\end{figure}

In this study, we have used both numerical and asymptotic approaches to analyse the mean flow
corrections generated by the nonlinear self-interaction of a harmonic solution in a rotating, librating spherical shell.
We have focused on the case where the inner core librates at a frequency $\widehat{\omega} = \sqrt{2}\widehat{\Omega}$, for which the harmonic solution
exhibits a simple structure,  primarily composed of critical point beams propagating along a closed rectangular periodic path.
Using the asymptotic structure of the harmonic solution  obtained in \citet{he2022}, we have obtained expressions for the dominant mean flow corrections in the limit of small Ekman numbers.
These theoretical predictions have been compared with numerical results obtained for Ekman  numbers ranging from $10^{-8}$ to $10^{-10}$, providing strong validation for both the numerical and asymptotic methods.
In addition, scaling laws for weaker mean flow bands have  been derived and compared against numerical data.
This analysis  has allowed us to obtain a comprehensive  picture of the mean flow correction structure, summarized for the three velocity components in figures~\ref{fig: scaling-v0phi},~\ref{fig: scaling-v0z} and~\ref{fig: scaling-v0r}.

The main findings are as follows.
We have shown that the largest mean flow corrections are localized  in the regions where two beams propagating in different directions intersect. In these localized overlap regions,
the three velocity components share the same scaling. The largest correction, of order  $E^{-1/2}$, occurs in the local region $P_1$ where the critical beam reaches the rotation axis. Significant corrections of order $E^{-1/6}$ have also been found in regions $P_3$ and $P_7$,
where the critical beam reflects on the outer boundary, and in  region $P_4$, where it intersects its symmetric counterpart on the equator. Weaker contributions, scaling as $E^{0}$ have  been observed in regions $P_5$ and $P_6$ where the main critical beam interacts with the wider and weaker secondary beam propagating between the critical points on the inner and outer core.

We have further demonstrated that these strong localized interactions could serve as sources of weaker mean flow bands aligned with  the rotation axis.
From  regions $P_3$ and $P_7$ on the outer boundary,  we have shown that  mean flow bands are generated, characterized by  axial and azimuthal velocities of order  $E^{0}$ and a radial velocity of order $E^{1/3}$, thus
confirming
the viscous generation mechanism proposed by \citet{ledizes2020}.  When the local interaction occurs away from the boundary, as in regions $P_1$ and $P_4$, the resulting   mean flow bands are relatively weaker.
In all of these bands, each of width $E^{1/3}$,  the velocity and pressure fields have a complex axial structure and therefore do not form  Taylor-Proudman columns.
In contrast, broader mean flow bands are  generated from the critical point region $P_2$ and $P_6$ located on the inner and outer cores. These bands  behave as Taylor-Proudman columns,
characterised by a dominant azimuthal velocity component that is independent of the axial coordinate, accompanied by much  weaker axial and radial velocity components.

In this study, we have focused on the mean flow correction. However, double-harmonic corrections, scaling as the square of the harmonic solution, are also expected.
For the frequency  considered, $2\omega$ lies outside the frequency range of inertial waves. As a result, double-harmonic corrections are  not expected to propagate.
Nevertheless, as explained in \citet{ledizes2020}, such corrections are still generated within the interaction regions. We expect them to exhibit the same scaling as  the mean flow corrections  in
the local regions $P_\beta$, but without the emission of beams from these regions.

In the present study, we have considered a viscous libration forcing of the inner core, which gives rise to an harmonic response of order $E^{1/12}$. For a larger harmonic response,  such as that obtained with inviscid forcing, larger mean flow corrections are expected.
\citet{he2025} considered the same geometry but with a different forcing, corresponding to an inner core vertical oscillation. When the same frequency is used, they showed that
a similar harmonic solution is obtained, concentrated along the same rectangular critical ray pattern,  but characterized by a different similarity index, $m=1/2$, and a larger amplitude scaling as $E^{-1/6}$.
A similar analysis can be applied to this solution, which has an amplitude $E^{-1/4}$ larger than the present harmonic solution.
We expect analogous results for the mean flow corrections generated by the interaction of the critical beam with itself, specifically within the local interaction regions $P_\beta$, as well as in the bands originating from these points.
However,  since the mean flow corrections scale with the square of the harmonic solution amplitude, all such corrections
would be amplified by a  factor of $E^{-1/2}$ in that case.  In particular, this leads to amplitude scalings of $E^{-3/2}$ in $P_1$,  $E^{-2/3}$ in $P_3$, $P_4$ and $P_7$, and  $E^{-1/2}$ in the bands
originating from $P_3$, $P_6$ and $P_7$.

Our results give some theoretical grounds to previously observed results in the literature.
For example, the fact that differential rotation is preferentially driven at locations where wave beams reflect on the boundaires has been observed in the tidally-driven zonal flows of \citet{favier2014}, albeit at much higher Ekman number that those discussed here and for different forcing frequencies.
Additionnally, tentative scaling for the volume-averaged energy of the differential rotation were reported in \citet{tilgner2007} and \citet{favier2014}, with exponents ranging from $E^{-1/2}$ to $E^{-3/2}$ (see also the scaling of $E^{-3/10}$ for the azimuthal velocity observed experimentally by \citet{morize2010} but for a full sphere, less relevant to our particular spherical shell).
Since these authors considered tidally-dirven flows, their scaling are to be compared with our prediction for an inviscid forcing.
From our local analysis, we expect scaling for the volume-averaged energy of the differential rotation of $E^{-2}$ for the local region around $P_1$ and $E^{-2/3}$ for the local regions around $P_3$, $P_4$ and $P_7$.
The volume-averaged energy associated with the bands emanating from $P_3$, $P_6$ and $P_7$ are predicted to scale as $E^{-2/3}$.
Our predictions are compatible with existing scaling, in particular with the fact that we expect intense zonal flows in the limit of vanishing Ekman number.
Note however that our approach remains valid only in the weakly non-linear limit but clearly disentangles the different zonal flow contributions which was not the case of the volume-averaged approach used in \cite{tilgner2007} and \cite{favier2014}.
Note finally that the strongest response on the rotation axis around $P_1$ might be connected to the focusing effect discussed by \cite{shmakova2021} and \cite{liu2022} in the case of an oscillating torus.
The fact that both the harmonic solution and the mean flow corrections reach their largest amplitudes near $P_1$ allows us to consider the future development of a strongly nonlinear theory, in which the nonlinear effects are concentrated in the vicinity of that point.
At a qualitative level, our predicted localized corrections and axial bands agree with the experimental steady zonal flow patterns of \citet{subbotin2022,subbotin2023}, but we refrain  from making a quantitative comparison because the forcings and accessible Ekman numbers in those experiments differ significantly from the asymptotic regime considered in our study.

These scalings, for both inviscid and a viscous forcing, define the limits of validity for the weakly nonlinear approach employed in this study.
The requirement that the mean flow corrections remain small  implies that the forcing amplitude $\epsilon$ must be much smaller than $E^{1/4}$ in the viscous case, and much smaller than
$E^{3/4}$ in the inviscid case.
An even more stringent condition arises from requiring that  the  mean flow correction remains  smaller than the harmonic solution. 
 This leads to the constraint  $\epsilon \ll E^{5/12}$ for the viscous case,
 and $\epsilon \ll E^{7/6}$ in the inviscid case.
This condition also ensures that the gradient of the mean flow correction remains small, thereby guaranteeing that the propagation of the harmonic beam is not perturbed by the mean flow correction at leading order.

The weakly nonlinear solution may also lose its physical relevance if it becomes unstable. The harmonic solutions exhibit strong shear regions that are potentially unstable.
A crude estimate of the instability threshold associated with this shear is obtained by balancing the vorticity within each layer against the viscous damping of a perturbation with a characteristic wavelength comparable to the layer width. For the libration forcing, this yields the condition 
$\varepsilon >O(E^{1/2}) $ for  instabilities in the Stokes–Ekman layer on the inner sphere, 
and $\varepsilon >O(E^{7/12}) $ for instabilities in both the main internal shear layer of width $E^{1/3}$ and
the secondary internal shear layer of width $E^{1/6}$. 
These constraints are more restrictive than the condition of validity of the weakly nonlinear analysis. It would be valuable to obtain more precise estimates for the instability thresholds through a dedicated stability analysis of the harmonic solution or by means of direct numerical simulations.

It is worth emphasizing that the results  presented here are for a harmonic solution with a relatively simple structure, in which the number of beam crossings is limited.
For other forcing frequencies,  the harmonic response can be significantly more complex, involving  multiple reflections of the critical point beams and the possible formation
of wave attractors \citep{he2023}.
\begin{figure}
	\centering
	\includegraphics[width=\textwidth]{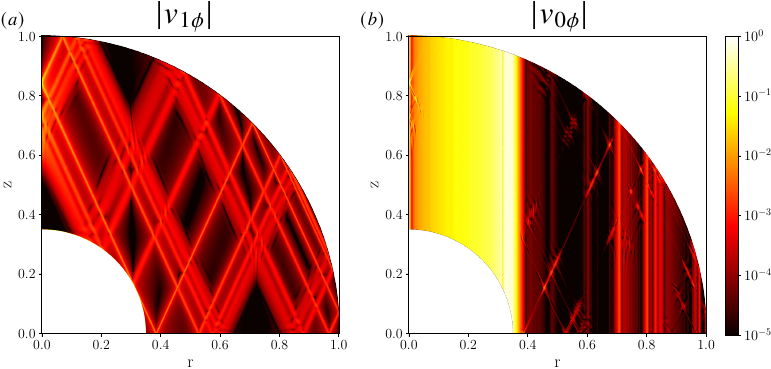}
	\caption{(a) Contours of amplitudes of harmonic linear azimuthal velocity $|v_{1\phi}|$. The blue lines are a critical line from the source point $S_{c}$. (b) Mean flow azimuthal velocity $|v_{0\phi}|$ at $E=10^{-9}$ and the libration frequency of the inner core is $\omega=0.8317$.}
	\label{fig: attractor}
\end{figure}
An illustration of such a harmonic response is shown in figure~\ref{fig: attractor}(a) for a libration frequency $\widehat{\omega} = 0.8317 \widehat{\Omega}$.
In figure~\ref{fig: attractor}(b), we present the azimuthal velocity contours of the mean flow correction obtained numerically from this harmonic response.
Despite the added complexity, many features of the mean flow corrections remain similar.
The dominant contribution arising from the boundary layer oscillation of the inner core is still present.
Localized contributions at  beam intersection points in the bulk are also clearly visible, as are bands originating from
locations where beams reflect on the boundaries.
For both the localized contributions and the emitted bands, a similar asymptotic analysis is expected to hold.  In fact, the structure of the solution may be simpler in this case,  as
the main beam is now a single critical point beam (with the same similarity index $m=5/4$ and amplitude scaling as $E^{1/12}$),  propagating in only one direction - unlike in our previous case, which involved an infinite sum of beams propagating in
both directions. As long as the analysis is restricted to regions away from the attractor, we therefore expect similar  localized contributions of order $E^{-1/6}$ at beam intersections, and
bands of width $O(E^{1/3})$,   with  velocity scaling as $E^0$, originating from reflection points on the outer boundary.

\backsection[Acknowledgements]{X. C. performed the numerical simulations, conducted the asymptotic
	analysis and the comparison between theory and numerics, created all the
	figures, and wrote the original draft.
	J. H. developed the numerical code.
	B. F. designed the study, supervised the numerical aspects, and
	reviewed/edited the manuscript.
	S. L. D. designed the study, supervised the theoretical aspects, and
	reviewed/edited the manuscript. The Centre de Calcul Intensif d’Aix-Marseille is acknowledged for granting access to its high-performance computing resources. This work was granted access to the HPC resources of IDRIS under the allocations A0140407543 and A0180407543 made by GENCI. The authors thank Prof. Yufeng Lin for providing the numerical simulation data for comparison.}

\backsection[Declaration of interests]{The authors report no conflict of interest.}

\backsection[Supplementary materials]{The directory stores the data, scaling parameters, and Jupyter notebook used for the validation of Ekman scaling at various cut positions (shown in figure~\ref{fig: cuts-pos}) from numerical computations at \url{https://www.cambridge.org/S0022112025109841/JFM-Notebooks/files/Figure_num_scaling/mean-num-
scaling.ipynb}. This validation corresponds to figure~\ref{fig: num-p3} and figures in section~\ref{sec: num-scalings}.}

\backsection[Author ORCIDs]{

	Xu Chang, https://orcid.org/0000-0002-5200-3022;

	Jiyang He, https://orcid.org/0000-0003-4176-1829;

	Benjamin Favier, https://orcid.org/0000-0002-1184-2989;

	St\'ephane  Le Diz\`es, https://orcid.org/0000-0001-6540-0433.}

\appendix
\section{Numerical convergence details}\label{sec: app-num}
\begin{figure}
	\centering
	\includegraphics[width=0.95\textwidth]{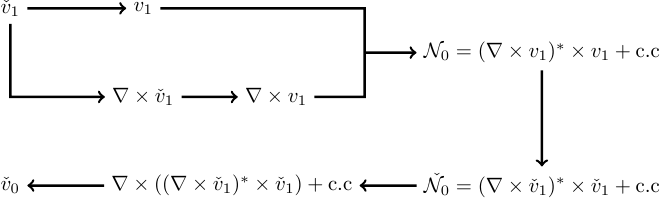}
	\caption{Schematic of the pseudo-spectral method workflow.}
	\label{fig: num-map}
\end{figure}

The pseudo-spectral method workflow is shown in figure~\ref{fig: num-map}. Based on the linear velocity in spectral space $\check{v}_1$ of the harmonic solution, this paper calculates the steady Reynolds stress through the pseudo-spectral method and then obtains the mean flow. For the angular components, we employ the SHTns package to efficiently transform spherical harmonic coefficients to physical space \citep{schaeffer2013}.

We tested two dealiasing approaches: traditional $2/3$ truncation for the spherical harmonic coefficients and the SHTns package's anti-aliasing capability for angular dimensions \citep{schaeffer2013}. No dealiasing was needed in the radial direction as we used Chebyshev differentiation matrices in physical space. Since both methods yielded identical results to simulations without dealiasing at our resolution, we present all results without applying dealiasing operations.

\begin{figure}
	\centering
	\includegraphics[width=0.95\textwidth]{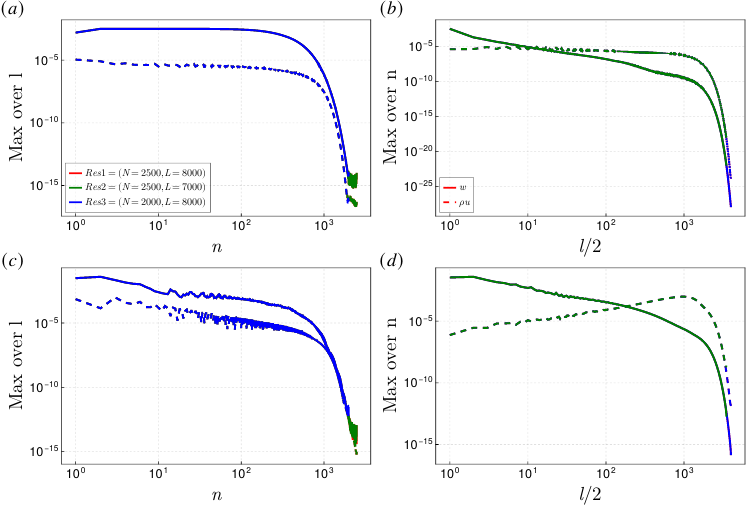}
	\caption{Spectra compare of $(a)$  linear Chebyshev coefficients, $(b)$ linear spherical harmonic components, $(c)$  nonlinear Chebyshev coefficients, and $(d)$ nonlinear spherical harmonic components at $E=10^{-10}$ with three different resolutions $Res1 = (2500,8000)$, $Res2=(2500,7000)$, and $Res3=(2000,8000)$, red, green and blue lines respectively. The toroidal component ($w$ solid line) and the radial component ($\rho u$ dashed line) have been separated. For each $l$ and $n$, the maximum value over the other spectral component is chosen.}
	\label{fig: spectra}
\end{figure}

The convergence of the spectral codes with various resolutions is tested by the spectra of the Chebyshev coefficients and the spherical harmonic components, as in \citet{rieutord1997}. Figure~\ref{fig: spectra} shows the spectra at the smallest Ekman number $E=10^{-10}$.
We have verified that for all Ekman numbers, the resolutions shown in table~\ref{tab: resolution} ensure that the ratio $Es_{\text{min}}/Es_{\text{max}}$ is less than $10^{-8}$, where $Es_{\text{min}}$ and $Es_{\text{max}}$ are the minimum and maximum values of the energy spectrum, respectively. This ensures that all of the length scales are properly resolved down to small dissipative structures, as is attested by the exponential cut-off observed at high wave numbers in figure~\ref{fig: spectra}.
\begin{table}
	\centering
	\def ~{\hphantom{0}}
	\begin{tabular}{lccc}
		Ekman number & Chebyshev grid resolution $N$ & Spherical harmonic grid resolution $L$ \\
		\hline
		$E=10^{-6}$  & 300                           & 600                                    \\
		$E=10^{-7}$  & 300                           & 900                                    \\
		$E=10^{-8}$  & 900                           & 3000                                   \\
		$E=10^{-9}$  & 900                           & 3500                                   \\
		$E=10^{-10}$ & 2500                          & 8000                                   \\
	\end{tabular}
	\caption{Resolutions of different Ekman numbers}\label{tab: resolution}
\end{table}

\section{Parameters for the harmonic solution}\label{sec: app-parameters}
The parameters defining the harmonic solution along the main rectangular circuit  are given in tables ~\ref{tab: north-path} and ~\ref{tab: south-path} for the northward and southward beam, respectively.
\begin{table}
	\begin{center}
		\def\arraystretch{1.3}
		\begin{tabular}{ccccc}
			\text{Critical line} & $x_{\parallel}^{N}$                                                              & $x_{\perp}^{N}$                                         & $\tilde{C}_{0}^{N}$                     & $L^{N}$ \\ [4pt]
			$S_{c}\to P_{1}$     & $\displaystyle\frac{-r+z}{\sqrt{2}}$                                             & $\displaystyle\frac{r+z-\sqrt{2}\eta}{\sqrt{2}}$        & $\tilde{C}_{0}^{N}$                     & $L$     \\ [8pt]

			$P_{1}\to P_{3}$     & $\displaystyle l_{1}+\frac{r+z-\sqrt{2}\eta}{\sqrt{2}}$                          & $\displaystyle\frac{-r+z-\sqrt{2}\eta}{\sqrt{2}}$       & $\mathrm{i}\tilde{C}_{0}^{N}$           & $L$     \\ [8pt]

			$P_{3}\to P_{7}$     & $\displaystyle(l_{1}+l_{2})K^{3}+\frac{r-z+\sqrt{2}\eta}{\sqrt{2}}$              & $\displaystyle\frac{-r-z+\sqrt{2-2\eta^{2}}}{\sqrt{2}}$ & $\mathrm{i}K^{-1 / 4}\tilde{C}_{0}^{N}$ & $K^3L$  \\ [8pt]

			$P_{7}\to P_{4}$     & $\displaystyle l_{1}+l_{2}+l_{3}K^{-3}+\frac{-r-z+\sqrt{2-2\eta^{2}}}{\sqrt{2}}$ & $\displaystyle\frac{r-z-\sqrt{2}\eta}{\sqrt{2}}$        & $\mathrm{i}\tilde{C}_{0}^{N}$           & $L$     \\ [8pt]

			$P_{4}\to S_{c}$     & $\displaystyle l_{1}+l_{2}+l_{3}K^{-3}+l_{4}+\frac{-r+z+\sqrt{2}\eta}{\sqrt{2}}$ & $\displaystyle\frac{r+z-\sqrt{2}\eta}{\sqrt{2}}$        & $\mathrm{i}\tilde{C}_{0}^{N} $          & $L$
		\end{tabular}
		\caption{Northward path coordinates, amplitudes and propagation distances within one cycle.}
		\label{tab: north-path}
	\end{center}
\end{table}
\begin{table}
	\begin{center}
		\def\arraystretch{1.3}
		\begin{tabular}{ccccc}
			\text{Critical line} & $x_{\parallel}^{S}$                                                              & $x_{\perp}^{S}$                                        & $\tilde{C}_{0}^{S}$           & $L^{S}$ \\ [4pt]
			$S_{c}\to P_{4}$     & $\displaystyle\frac{-r-z+\sqrt{2}\eta}{\sqrt{2}}$                                & $\displaystyle\frac{-r-z+\sqrt{2}\eta}{\sqrt{2}}$      & $\tilde{C}_{0}^{S}$           & $L$     \\ [8pt]

			$P_{4}\to P_{7}$     & $\displaystyle l_{5}+\frac{r+z-\sqrt{2}\eta}{\sqrt{2}}$                          & $\displaystyle\frac{-r+z+\sqrt{2}\eta}{\sqrt{2}}$      & $\tilde{C}_{0}^{S}$           & $L$     \\ [8pt]

			$P_{7}\to P_{3}$     & $\displaystyle(l_{5}+l_{4})K^{3}+\frac{-r+z+\sqrt{2}\eta}{\sqrt{2}}$             & $\displaystyle\frac{r+z-\sqrt{2-2\eta^{2}}}{\sqrt{2}}$ & $K^{-1 / 4}\tilde{C}_{0}^{S}$ & $K^3L$  \\ [8pt]

			$P_{3}\to P_{1}$     & $\displaystyle l_{5}+l_{4}+l_{3}K^{-3}+\frac{-r-z+\sqrt{2-2\eta^{2}}}{\sqrt{2}}$ & $\displaystyle\frac{r-z+\sqrt{2}\eta}{\sqrt{2}}$       & $\tilde{C}_{0}^{S}$           & $L$     \\ [8pt]

			$P_{1}\to S_{c}$     & $\displaystyle l_{5}+l_{4}+l_{3}K^{-3}+l_{2}+\frac{r-z+\sqrt{2}\eta}{\sqrt{2}}$  & $\displaystyle\frac{-r-z+\sqrt{2}\eta}{\sqrt{2}}$      & $\mathrm{i}\tilde{C}_{0}^{S}$ & $L$
		\end{tabular}
		\caption{Southward path coordinates, amplitudes and propagation distances within one cycle.}
		\label{tab: south-path}
	\end{center}
\end{table}
They use the following quantities
\be
L=l_1+l_2+l_3K^{-3}+l_4+l_5
\label{eq: L}
\ee
with
\begin{equation}
	l_{1}=\eta,\quad l_{2}=\sqrt{ 1-\eta^{2} }-\eta,\quad l_{2}=2\eta,\quad l_{4}=l_{2},\quad l_{5}=l_{1},
\end{equation}
and
\begin{equation}
	K=\frac{\sin(\alpha+\pi/4)}{\sin(\alpha-\pi/4)} .
\end{equation}

\section{Mean flow bands}
\label{app:band}

In this section, we analyse the structure of the mean flow bands that are created within the bulk from the local interaction regions $P_\beta$.
These bands have the particularity to be present in fluid regions where the Reynolds stress is very small.
The corresponding velocity field is therefore expected to satisfy  homogeneous equations.

\subsection{Bands issued from $P_3$, $P_4$ and $P_7$}

We first consider bands generated from the local regions $P_3$, $P_4$ and $P_7$. These regions have a width of order $E^{1/3}$ and we can expect the bands to have  a same width. It is thus natural
to introduce the local radial variable $\tilde{r}=E^{-1/3}(r-r_{P_\beta})$ and to use the following  ansatz for the velocity field:
\begin{equation}
	{\bf v}_0 = (E^{1/3} v_{0r}, v_{0\phi}, v_{0z}, E^{1/3} p_0) (\tilde{r},z) .
	\label{exp:v0bulk}
\end{equation}
The velocity field then satisfies the following set of equations
\bsea
-2 v_{0\phi} + \frac{\partial p_0}{\partial \tilde{r}} =0 ,\\
2 v_{0r} -  \frac{\partial^2 v_{0\phi}}{\partial \tilde{r}^2} =0 ,\\
\frac{\partial p_0}{\partial z} -  \frac{\partial^2 v_{0z}}{\partial \tilde{r}^2} =0 , \\
\frac{\partial v_{0r}}{\partial \tilde{r}} + \frac{\partial v_{0z}}{\partial z} =0 .
\esea
For $P_3$ and $P_7$, the solution should be valid from $-z_{P_\beta}$ to $z_{P_\beta}$ with a condition of anti-symmetry on the axial velocity component
with respect to the equatorial plane.
This implies that the general solution for the bands issued from $P_3$ and $P_7$ takes the form
\bsea
&&v_{0r} = -\frac{\mathrm{i}}{4} \int_{-\infty}^{+\infty} A(k) k^3\cosh(k^3z/2)e^{\mathrm{i}k\tilde{r}} dk , \\
&&v_{0\phi} = -\frac{\mathrm{i}}{2} \int_{-\infty}^{+\infty} A(k) k \cosh(k^3z/2)e^{\mathrm{i}k\tilde{r}} dk, \\
&&v_{0z} =-\frac{1}{2} \int_{-\infty}^{+\infty} A(k) k \sinh(k^3z/2)e^{\mathrm{i}k\tilde{r}} dk, \\
&&p_{0} =  \int_{-\infty}^{+\infty} A(k)  \cosh(k^3z/2)e^{\mathrm{i}k\tilde{r}} dk.
\esea
where $A(k)$ is a function determined by the boundary condition at $z=z_{P_\beta}$.
In particular,  $A(k)$ is set by the Ekman pumping generated
at the boundary close to $P_\beta$.
We have seen that, at the order $E^{-1/6}$, the Ekman pumping vanishes.
This explains why the order of the band is smaller than $E^{-1/6}$.
As shown in \citet{ledizes2020}, the Ekman pumping first appears at the order $E^{0}$.
If we denote by $w_{\infty} (\tilde{r}) = \tilde{v}_{0z}(\tilde{r},\tilde{z}=-\infty)$ the Ekman puming at this order,  the function  $A(k)$ is just obtained by
the condition $v_{0z}(\tilde{r},z_{P_\beta}) = w_{\infty} (\tilde{r})$, which leads to
\be
A(k) = -\frac{2\hat{w}_\infty}{k\sinh(k^3z_{P_\beta}/2)} ~,
\ee
where $\hat{w}_\infty$ is the Fourier transform of $w_\infty$.

For the band issued from the point $P_4$ on the equatorial plane, the solution must satisfy
non-penetration conditions at the ends of the band, corresponding to $z=\pm z_{4}=\pm \sqrt{1-2\eta^2}$.
This leads to the following form for $z>0$:
\bsea
&&v_{0r} = -\frac{\mathrm{i}}{4} \int_{-\infty}^{+\infty} A(k) k^3\cosh(k^3(z-z_{4})/2)e^{\mathrm{i}k\tilde{r}} dk , \\
&&v_{0\phi} = -\frac{\mathrm{i}}{2} \int_{-\infty}^{+\infty} A(k) k \cosh(k^3(z-z_{4})/2)e^{\mathrm{i}k\tilde{r}} dk, \\
&&v_{0z} =-\frac{1}{2} \int_{-\infty}^{+\infty} A(k) k \sinh(k^3(z-z_{4})/2)e^{\mathrm{i}k\tilde{r}} dk, \\
&&p_{0} =  \int_{-\infty}^{+\infty} A(k)  \cosh(k^3(z-z_{4})/2)e^{\mathrm{i}k\tilde{r}} dk.
\esea
The amplitude $A(k)$ is now prescribed by the axial flux generated from the Reynolds stress close to $z=0$ around $P_4$.
We have seen that it is zero at the order $E^{-1/6}$.
We further claim that it also vanishes at the orders $E^{-1/12}$ and $E^{0}$, due to  the similar form of the Reynolds stress at these orders.
Indeed, as shown in section~\ref{sec:higher}, up to $O(E^{1/3})$, the viscous corrections to the leading-order harmonic solution has a similar structure to the main beam. They also satisfy
the two key properties -\eqref{eq: v1phi2} and \eqref{eq: v1r}- that were used to derive the mean flow correction expression \eqref{exp:v0gen}, which notably exhibits no axial flux.

The first correction that  generates a non-zero axial flux from $P_4$ is therefore expected to appear at a magnitude $E^{1/3}$ smaller than the dominant term - that is,  at the order $E^{1/6}$.
If this flux is denoted as $E^{1/6} w_\infty(\tilde{r})$, the function $A(k)$ is obtained by enforcing the condition $v_{0z}(\tilde{r},z=0) = E^{1/6}w_\infty(\tilde{r})$, leading to
\be
A(k) = E^{1/6}\frac{2\hat{w}_\infty}{k\sinh(k^3z_{Q_4}/2)} ~.
\ee
We therefore expect the  band generated from $P_4$ to be of order $E^{1/6}$ smaller than those originating from $P_3$ and $P_7$.

It is worth emphasizing that the velocity field in the bands issued from the three points $P_3$, $P_4$ and $P_7$ depends on the axial
coordinate $z$. These bands are therefore not Taylor-Proudman columns, in which axial and azimuthal velocity components are invariant along the rotation axis.
Here, due the smallness of the radial scale, the viscous effects responsible for axial variations become observable over the O(1) axial extent of the column.

\subsection{Band issued from $P_1$}

For the band originating from the point $P_1$, the analysis differs slightly due to the presence of the cylindrical singularity.
We still introduce the local variable $\tilde{r}=E^{-1/3}r$,  and adopt the same ansatz \eqref{exp:v0bulk} for the mean flow corrections.
The governing equations become:
\bsea
-2 v_{0\phi} + \frac{\partial p_0}{\partial \tilde{r}} =0 ,\\
2 v_{0r} -  \left(\frac{\partial^2}{\partial \tilde{r}^2}   + \frac{1}{\tilde{r}}\frac{\partial}{\partial \tilde{r}} -\frac{1}{\tilde{r}^2}\right)v_{0\phi} =0 ,\\
\frac{\partial p_0}{\partial z} - \left(\frac{\partial^2}{\partial \tilde{r}^2}   + \frac{1}{\tilde{r}}\frac{\partial}{\partial \tilde{r}}\right) v_{0z} =0 , \\
\frac{\partial v_{0r}}{\partial \tilde{r}}+ \frac{v_{0r}}{\tilde{r}} + \frac{\partial v_{0z}}{\partial z} =0 .
\esea
On either side of $P_1$, we obtain solutions that satisfy the non-penetration condition at $z=z^+_1=1$ and at $z=z_1^-=\eta$, respectively. These solutions take  the following form:
\bsea
&&v_{0r}^\pm = \frac{1}{4} \int_{-\infty}^{+\infty} A^{\pm}(k) k^3\cosh(k^3(z-z^{\pm}_1)/2)J_1(k\tilde{r}) dk , \\
&&v_{0\phi}^\pm = -\frac{1}{2} \int_{-\infty}^{+\infty} A^{\pm}(k) k \cosh(k^3(z-z^{\pm}_1)/2)J_1(k\tilde{r})dk, \\
&&v_{0z}^\pm =-\frac{1}{2} \int_{-\infty}^{+\infty} A^{\pm}(k) k \sinh(k^3(z-z^{\pm}_1)/2)J_0(k\tilde{r}) dk, \\
&&p_{0}^\pm =  \int_{-\infty}^{+\infty} A^{\pm}(k)  \cosh(k^3(z-z^{\pm}_1)/2)J_0(k\tilde{r}) dk,
\esea
where the superscript $+$ refers to the region $z_{P_1} < z \leq 1$, and the superscript $-$ refers to $\eta \leq z < z_{P_1}$.
The two functions $A^{\pm}(k)$ are obtained by the conditions of matching with the local solution close to $P_1$.
As previously shown, at the order $E^{-1/2}$, the local solution is confined to the vicinity of  $P_1$, and thus does not generate any mean flow bands.
For the same reasons as in the case of $P_4$, the corrections at the orders $E^{-5/12}$ and $E^{-1/3}$ are also localized and do not induce any non-local band structure.
The first non-localized contribution is expected to arise at the order $E^{-1/6}$. The functions $A^{\pm}$ can be related
to the jumps $E^{-1/6}\delta \hat{v}$ and $E^{-1/6}\delta \hat{w}$ of the azimuthal and axial velocity  across the local region around $P_1$
by the following relations
\bsea
v_{0\phi}^+ (\tilde{r}, z=z_{P_1}) - v_{0\phi}^- (\tilde{r}, z=z_{P_1}) =E^{-1/6} \delta \hat{v}(\tilde{r}) ,\\
v_{0z}^+ (\tilde{r}, z=z_{P_1}) - v_{0z}^- (\tilde{r}, z=z_{P_1}) = E^{-1/6}\delta \hat{w}(\tilde{r}) .
\esea

As with the bands originating from $P_3$, $P_4$ and $P_7$, the band emanating from $P_1$ also
exhibit axial dependence.

\subsection{Bands issued from $P_2$ and $P_6$}

The bands issued from the local regions $P_2$ and $P_6$  differ from those previously discussed, as their radial width is larger than $E^{1/3}$.
As a consequence, the azimuthal velocity can no longer depend on the axial coordinate $z$. These bands are then Taylor-Proudman columns.

For $P_6$, the appropriate radial variable  and velocity ansatz are $\breve{r} = (r-r_{P_6})/E^{1/6}$
and
\begin{equation}
	{\bf v}_0 = (E^{2/3} v_{0r}, v_{0\phi}, E^{1/2} v_{0z}, E^{1/6} p_0) (\breve{r},z) .
	\label{exp:v0bulkP6}
\end{equation}
This leads to the following governing equations:
\bsea
-2 v_{0\phi} + \frac{\partial p_0}{\partial \breve{r}} =0 ,\\
2 v_{0r} -  \frac{\partial^2 v_{0\phi} }{\partial \breve{r}^2}    =0 ,\\
\frac{\partial p_0}{\partial z} =0 , \\
\frac{\partial v_{0r}}{\partial \breve{r}} + \frac{\partial v_{0z}}{\partial z} =0 .
\label{eq:v0P6}
\esea
This ansatz and these equations above show that an azimuthal velocity of order 1 in the bulk can only be compatible with an axial velocity  of order $E^{1/2}$.
In particular,  (\ref{eq:v0P6}b,d)  imply that the axial velocity, which has to be antisymmetric with respect to the equatorial plane,
must satisfy
\be
v_{0z} = -z \frac{\partial v_{0r}}{\partial \breve{r}} = -\frac{z}{2} \frac{\partial^3 v_{0\phi}}{\partial \breve{r}^3},
\label{eq:v0zP6}
\ee
for a given azimuthal velocity profile $v_{0\phi}(\breve{r})$.
However,  the $O(1)$ particular solution that we computed in \S \ref{sec: P5} is expected to induce  Ekman pumping of order $E^{1/3}$.
Similarly, an $O(1)$ azimuthal velocity  in the bulk also generates Ekman pumping of the same order.
These two contributions must cancel each other to ensure compatibility with the axial flow in the bulk. This  matching condition prescribes
the function $v_{0\phi}(\breve{r})$, from which the full velocity and pressure field in the band is entierely determined.

The case of $P_2$  is more complex, as both $E^{1/5}$ and $E^{1/6}$ radial scales are, a priori, possible.
Assuming the  $E^{1/5}$ scale dominates, we introduce
$\bar{r} = (r-r_{P_2})/E^{1/5}$ and the ansatz
\begin{equation}
	{\bf v}_0 = (E^{3/5} v_{0r}, v_{0\phi}, E^{2/5} v_{0z}, E^{1/5} p_0) (\bar{r},z) .
	\label{exp:v0bulkP2}
\end{equation}
The resulting equations are analogous to those in \eqref{eq:v0P6} with $\breve{r}$ replaced by $\bar{r}$.
As in the case of $P_6$, the leading-order azimuthal velocity component must be independent of $z$.
However, the axial velocity may now either  be linear in $z$ with the scaling prescribed by the ansatz \eqref{exp:v0bulkP2}, or  independent of $z$ and larger in magnitude.
It is important to note  that the bulk solution with an $O(1)$ azimuthal velocity is expected to generate  $O(E^{3/10})$ Ekman pumping at both the inner and outer cores.
These Ekman pumpings differ on each
boundary and  are larger than the possible linear axial flow in the bulk. Therefore, the azimuthal velocity in bulk  must be chosen such that
the Ekman pumping generated in $P_2$, by the solution forced by the Reynolds stress,  exactly cancels  the Ekman pumping  produced by the bulk solution at the outer boundary.
This cancellation allows for a $z$-independent axial flow in the bulk at leading order.
Although evaluating the Ekman pumping precisely is challenging,  the above reasoning enables us to deduce the relative orders of the velocity components.
For instance,  if the Ekman pumping in $P_2$ is $O(E^{1/5})$, then, in the bulk, we expect: $v_{0\phi} = O(E^{-1/10})$,  $v_{0z}=O(E^{1/5})$, and $v_{0r}=O(E^{1/2})$. A weak axial dependence of the axial velocity is
also expected at the order $E^{3/10}$ and given by the analogue of equation \eqref{eq:v0zP6} for this band.

To conclude this section, we briefly comment on  the band originating from $P_5$.
As in the case of $P_1$, such a band is generated by the jumps of $v_{0\phi}$ and $v_{0z}$ across the local region $P_5$.
Due to the different scales of the two interacting beams, these jumps are expected to be of order $E^{2/3}$. The jump of $v_{0z}$  must be compensated by
an Ekman pumping of the same order at the outer boundary. This requires an azimuthal velocity in the band of order $E^{1/3}$. This velocity amplitude  is
comparable to the background azimuthal velocity in the bulk, which explains why no distinct band originating from $P_5$ is visible in figures~\ref{fig: contourmap}
and~\ref{fig: bulk-compare}(b).

\bibliographystyle{jfm}
\bibliography{jfm}

\end{document}